\documentclass[11pt]{article}

\usepackage{amsmath, amsfonts, amssymb}
\usepackage{accents}
\usepackage{color}
\usepackage{mathrsfs}
\usepackage{graphicx}
\usepackage{multirow}
\usepackage{subfigure}
\usepackage[hidelinks]{hyperref}
\usepackage{url}
\usepackage{fullpage}

\usepackage[normalem]{ulem}

\usepackage{tikz}
\usetikzlibrary{patterns}
\usepackage{subfigure}

\usepackage[numbers,sort&compress]{natbib}
\usepackage{authblk}

\newcommand\nc{\newcommand}
\nc\pa{\partial}
\nc\pad[2]{\frac{\pa #1}{\pa #2}}
\nc\padd[2]{\frac{\pa^2 #1}{\pa
{#2}^2}}
\nc\nd[2]{\frac{\text{d} #1}{\text{d} #2}}
\nc\ndd[2]{\frac{d^2 #1}{d {#2}^2}}
\nc\pat[2]{\frac{D #1}{D
#2}}
\nc{\ii}{\text{\textbf{i}}}
\nc\ra{\rightarrow} \nc\Ra{\Rightarrow}
\nc{\ud}{\mathrm{d}}

\newcommand{\Kn}{\text{Kn}}
\newcommand{\eff}{\text{eff}}
\newcommand{\kin}{\text{kin}}
\newcommand{\col}{\text{col}}

\newcommand{\rev}[1]{#1}

\newenvironment{subeq}[1]{
\begin{subequations}
#1
\end{subequations}
}

\title{Effective thermal conductivity of rectangular nanowires based on phonon hydrodynamics\footnote{Published in Int.~J.~Heat Mass Transfer: \url{https://doi.org/10.1016/j.ijheatmasstransfer.2018.05.096}}}

\date{}

\author[1,2]{M.~Calvo-Schwarzw\"alder}
\author[1]{M.~G.~Hennessy\footnote{Corresponding author: {\tt mhennessy@crm.cat}}}
\author[3]{P.~Torres}
\author[1,2]{T.~G.~Myers}
\author[3]{F.~X.~Alvarez}

\affil[1]{Centre de Recerca Matem\`{a}tica, Campus de Bellaterra, Edifici C, 08193 Bellaterra, Barcelona, Spain}
\affil[2]{Departament de Matem\`{a}tiques, Universitat Polit\`{e}cnica de Catalunya, Barcelona, Spain}
\affil[3]{Departament de F\'isica, Universitat Aut\`onoma de Barcelona, 08193 Bellaterra, Spain}

\begin{document}

\maketitle

\begin{abstract}
A mathematical model is presented for  thermal transport in nanowires with rectangular cross sections. Expressions for the effective thermal conductivity of the nanowire across a range of temperatures and cross-sectional aspect ratios are obtained by solving the Guyer--Krumhansl hydrodynamic equation for the thermal flux with a slip boundary condition. Our results show that square nanowires transport thermal energy more efficiently than rectangular nanowires due to optimal separation between the boundaries. However, circular nanowires are found to be even more efficient than square nanowires due to the lack of corners in the cross section, which locally reduce the thermal flux and inhibit the conduction of heat. By using a temperature-dependent slip coefficient, we show that the model is able to accurately capture experimental data of the effective thermal conductivity obtained from Si nanowires, \rev{demonstrating that phonon hydrodynamics is a powerful framework that can be applied in nanosystems even at room temperature.
\\[1em]
} 
\end{abstract}


\section{Introduction}
Advances in manufacturing processes have brought us to the stage where reliable nanoscale devices are now commonplace. However, in most current and predicted applications of nanostructures, there is a strong concern over the management of heat \cite{Cahill2003}. One specific issue with heat removal is the dramatic decrease of the thermal conductivity at the nanoscale in comparison to the bulk value \cite{Inyushkin2004,Li2003,Ashenghi1997,Liu2004}. This decrease can be rationalised in terms of the manner by which thermal energy is transported across the macroscale and nanoscale. At the macroscale, heat transfer is a diffusive process driven by frequent collisions between thermal energy carriers known as phonons. In contrast, the transport of thermal energy across the nanoscale is a ballistic process driven by infrequent collisions between phonons. As the size of a device approaches that of the phonon mean free path, the phonons become more likely to collide with a boundary than with each other. This results in a conductivity which is more strongly influenced by the scattering dynamics at the boundary than collisions in the bulk. Consequently, thermal energy is transported less efficiently across the nanosystem, yielding a decrease in the thermal conductivity from the bulk value of the material.

Developing mathematical models to aid in the understanding of heat flow has proved problematic due to the breakdown of Fourier's law at small time and length scales \cite{Hoogeboom-Pot2015,Johnson2013,Chang2008}. With the aim of accurately predicting the effective thermal conductivity (ETC) of nanosystems, a variety of theoretical models of nanoscale heat transport have been proposed, either based on micro- and mesoscopic approaches \cite{Callaway1959,Holland1963} or from a macroscopic point of view \cite{Calvo2018,Zhu2017,Alvarez2009,Sellitto2012,Alvarez2007,Ma2012,Dong2014,Tzou2011}. A popular approach is the phonon hydrodynamics model \cite{Calvo2018,Zhu2017,Alvarez2009,Sellitto2012,Alvarez2007,Ma2012,Dong2014}, which is based on the Guyer--Krumhansl (G-K) equation \cite{Guyer1966a,Guyer1966b}. \rev{This model was first presented to describe heat transfer in the so-called hydrodynamic regime, where phonon flow behaves as a rarefied gas due to the dominance of normal scattering, i.e., scattering which conserves quasi-momentum, over resistive scattering, which does not conserve phonon quasi-momentum \cite{Guyer1966a,Guyer1966b,Guo2015}. It was originally believed that the hydrodynamic transport regime only occurs at extremely low temperatures. However, recent \emph{ab initio} calculations have demonstrated that phonon hydrodynamics can be valid even at room temperature \cite{Guo2018,Lee2015,Ziabari2018}.} As with many other macroscopic models such as the thermomass model \cite{Tzou2011,Dong2011} or the equation of phonon radiative transfer \cite{Majumdar1993}, the G-K equation can derived from the Boltzmann transport equation (BTE) \cite{Boltzmann1872} or from the extended irreversible thermodynamics (EIT) framework \cite{Jou1996}. Thus, the G-K equation provides a link between microscopic (kinetic) and macroscopic (continuum) models. Another attractive feature of the phonon hydrodynamics model is that the governing equations are analogous to those seen in viscous fluid mechanics, which provides an intuitive conceptual framework for model development and interpretation.  The analogy between the phonon hydrodynamics and fluid dynamics models has prompted researchers to apply the well-known slip boundary condition to thermal transport  \cite{Calvo2018,Zhu2017,Alvarez2009,Sellitto2012}. \rev{In fact, it has been shown that this form of boundary condition naturally arises from the discrete BTE \cite{Xu2014} and from the EIT framework \cite{Guo2015}.} With the correct choice of slip length, this approach was shown to provide excellent agreement with experimental measurements of the ETC of silicon nanowires \cite{Calvo2018}.

Aside from circular nanowires, studies employing the slip boundary condition are mainly confined to two-dimensional thin-film geometries \cite{Zhu2017, Alvarez2009, Sellitto2012}. Although the reduced dimensionality of this geometry enables a simple, closed-form expression for the ETC to be obtained, it can only be applied to nanowires with extremely small cross-sectional aspect ratios. However, rectangular nanowires with aspect ratios as high as 0.64 have been reported in the literature \cite{Inyushkin2004}. Unlike the thin-film geometry, the cross section of a rectangular nanowire will have two additional boundaries, as well as four corners, that will be detrimental to thermal transport. Therefore, the geometry of the rectangular cross section will play a key role in determining the ETC of the nanowire and have practical consequences in terms of thermal regulation in nanodevices.

The purpose of the present study is to use the phonon hydrodynamic model with a slip boundary condition to calculate the ETC in a rectangular nanowire, with the aim of gaining an improved understanding of how finite cross-sectional geometries influence nanoscale heat transport. Of particular interest is determining the nanowire geometry that leads to the most efficient transport of heat. Theoretical predictions of the ETC are compared against experimentally measured values. We show that the model is able to accurately capture the experimental data across a wide range of temperatures.

The paper is organised as follows. The model and boundary conditions are presented in Sec.~\ref{sec:model} and expressions for the ETC are given in Sec.~\ref{sec:etc}. In Sec.~\ref{sec:res}, the modelling results are discussed and the theoretical predictions are compared against experimental data. The paper then concludes in Sec.~\ref{sec:conc}.


\section{Mathematical model}
\label{sec:model}

\begin{figure}
\centering
\begin{tikzpicture}[scale=0.66]
\fill [left color=red!80, right color=red!50,opacity=0.3] (-2,-1) rectangle (2,1);
\draw (-2,-1) rectangle (2,1);
\draw [dashed] (-2,-1) -- (1,2);
\draw (2,-1) -- (5,2);
\draw (-2,1) -- (1,4);
\draw (2,1) -- (5,4);
\draw [dashed] (1,2) rectangle (5,4);
\draw [<->] (2,-1.5) -- (5,1.5) node [midway,below] {$\ L^*_3$};
\draw [<->] (0,-1.5) -- (-2,-1.5) node [midway,below] {$L^*_1$};
\draw [<->] (-2.5,0) -- (-2.5,1) node [midway,left] {$L^*_2$};
\fill [left color=red!50, right color=blue!50,opacity=0.3] (2,-1)--(2,1)--(5,4)--(5,2);
\fill [left color=red!20, right color=blue!50,opacity=0.3] (-2,1)--(2,1)--(5,4)--(1,4);
\draw [->] (5,-.4) -- (6,-.4) node [midway,below] {$x^*$};
\draw [->] (5,-.4) -- (5,.6) node [midway,left] {$y^*$};
\draw [->] (5,-.4) -- (5.85,.45) node [right] {$z^*$};
\node at (-3.2,-.5) {$T_0^* + \Delta T^*$};
\node at (6,3) {$T_0^*$};
\end{tikzpicture}
\caption{The experimental setup consists of a rectangular nanowire of length $L^*_3$ and with a cross-section of dimensions $2L^*_1\times2L_2^*$. The heat flux $\textbf{Q}^*$ is induced by a constant temperature difference  $\Delta T^*>0$.}
\label{fig:rectangle}
\end{figure}
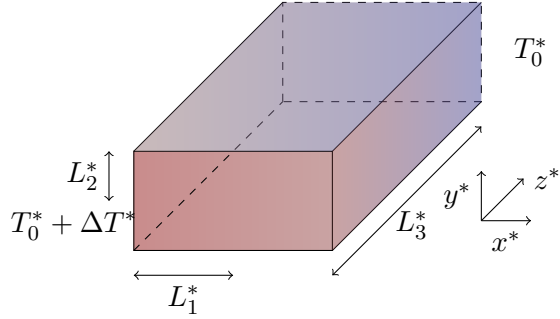

We model the transport of thermal energy across a long nanowire with rectangular cross section that is suspended in a vacuum; see Fig.~\ref{fig:rectangle}.  The length of the nanowire $L_3^*$ is assumed to be much greater than both the half-width $L_1^*$ and half-height $L_2^*$ of the cross section. The $^*$ notation is used to denote dimensional quantities.  Without loss of generality, the cross section can be taken to be wider than it is tall, $L_2^* / L_1^* < 1$.  Under these conditions, the cross-sectional aspect ratio $\phi = L_2^* / L_1^*$ and the longitudinal aspect ratio $\epsilon = L_2^* / L_3^*$ satisfy $\epsilon \ll \phi < 1$.  The transverse coordinates $x^*$ and $y^*$ denote points within a given cross section whereas the longitudinal coordinate $z^*$ describes distances along the length of the nanowire.  The origin of the cross section $(x^*,y^*) = (0,0)$ is chosen to coincide with the center of the rectangular face.  The transport of heat is driven by a longitudinal temperature gradient $\Delta T^* > 0$ that is imposed by fixing the temperature $T^*$ at the ends of the nanowire to be $T^* = T_0^* + \Delta T^*$ at $z^* = 0$ and $T^* = T_0^*$ at $z^* = L_3^*$.  The thermal flux is assumed to be symmetric about the $x^* = 0$ and $y^* = 0$ planes.

\subsection{Governing equations}
The mathematical model consists of an equation representing conservation of thermal energy and the G-K (or hydrodynamic) equation describing the evolution of the thermal flux.
Under the steady-state assumption, the governing equations are
\subeq{
\label{original_system}
\begin{align}
  \nabla\cdot\textbf{Q}^*&=0, \label{dim:div} \\
  \textbf{Q}^*&=-k^*\nabla T^*+\ell^{*2}\nabla^2\textbf{Q}^*, \label{dim:gk}
\end{align}
}
where $\textbf{Q}^*=u^*\hat{\mathbf{x}}^*+v^*\hat{\mathbf{y}}^*+w^*\hat{\mathbf{z}}^*$ is the thermal flux written in terms of Cartesian components, $k^*(T^*)$ is the bulk thermal conductivity, and $\ell^*(T^*)$ is a non-local length related to the bulk phonon mean free path (MFP), \emph{i.e.}, the mean distance between phonon-phonon collisions. \rev{In the original equation derived by Guyer and Krumhansl, the parameter $\ell^*$ represents the bulk MFP defined as $\ell^{*2}=(v_g^{*2}\tau_N^*\tau_R^*)/5$, where $v^*_g$ is the phonon group velocity and $\tau_N^*$ and $\tau_R^*$ are the normal and resistive mean free times, i.e., the mean times for normal and resistive scattering. Other researchers have proposed alternative definitions for $\ell^*$, such as the geometric mean of the bulk MFP and a local MFP, the latter of which decreases near a boundary \cite{Zhu2017}. We take $\ell^*$ to be the non-local length computed from the Kinetic Collective Model (KCM), which is an accurate approach for predicting the thermal conductivity in a number of materials \cite{Torres2017b}.  We refer to \ref{app:kcm} for a more detailed description of how $\ell^*$ is obtained from the KCM.}

For simplicity, the dependence of the parameters on the temperature will not be explicitly written unless required by the context. The second term on the right-hand side of \eqref{dim:gk} accounts for non-local effects, \emph{i.e.,} phonon collisions. The strong increase in the length scale $\ell^*$ as the temperature decreases can make non-local effects relevant in relatively large systems, including those with dimensions exceeding the nanoscale \cite{Jou1996, TorresThesis2017}.

The size-dependent effective thermal conductivity (ETC) of the nanowire $k_\text{eff}^*$ is defined as in terms of the mean thermal flux through a cross section $Q^*$ and the longitudinal temperature gradient $\pa T^*/\pa z^*$ as
\begin{align}\label{keff_dim}
k^*_{\text{eff}} = -\frac{Q^*}{\pa T^*/\pa z^*},\quad Q^* = \frac{1}{4L_1^*L_2^*} \int_{-L_2^*}^{L_2^*} \int_{-L_1^*}^{L_1^*} w^*\,\mathrm{d}x&^*\mathrm{d}y^*.
\end{align}
By construction, this is the only component that is not tangential to the cross-section. Therefore, to compute the effective thermal conductivity of a rectangular slab, we need to find only the normal component of the heat flux $w^*$. 

\subsection{Boundary conditions}\label{sec:bc}

The boundary conditions for the temperature at the endpoints of the nanowire are
\begin{subequations}
\begin{alignat}{2}
	T^*&=T_0^* + \Delta T^*, &\quad \text{ at } z^*&=0, \\
        T^*&=T_0^*, &\quad \text{ at } z^* &=L_3^*.
\end{alignat}
\end{subequations}
Furthermore, it is sufficient to solve the equations in the region $x^*,y^*\geq0$ due to the symmetry of the system.
The symmetry conditions at the interior boundaries ($x = 0$ or $y = 0$) are straightforward,
\begin{subequations}\label{boundary_conditions_flux}\begin{alignat}{7}
	&u^*=0,\qquad
	&&\pad{v^*}{x^*}=0,\qquad
	&&\pad{w^*}{x^*}=0,\qquad
	&& \text{at }x^*=0,\\
	&\pad{u^*}{y^*}=0,\qquad
	&&v^*=0,\qquad
	&&\pad{w^*}{y^*}=0,\qquad
	&& \text{at }y^*=0.
\end{alignat}
At the exterior boundaries of the nanowire ($x^* = L_1^*$ or $y^* = L_2^*$) we impose two types of boundary condition. The normal component of the flux is assumed to vanish to ensure that no energy is lost during the heating process. For components of the flux that are tangential to the surface, we impose slip conditions with a slip length $\ell^*_s$.  The boundary conditions are therefore given by
\begin{alignat}{7}
	&u^*=0,\qquad
	&&v^*=-\ell^*_s\pad{v^*}{x^*},\qquad
	&&w^*=-\ell^*_s\pad{w^*}{x^*},\qquad
	&& \text{at }x^*=L^*_1,\\
	&u^*=-\ell^*_s\pad{u^*}{y^*},\qquad
	&&v^*=0,\qquad
	&&w^*=-\ell^*_s\pad{w^*}{y^*},\qquad
	&& \text{at }y^*=L^*_2.
\end{alignat}\end{subequations}
\rev{Microscopically, slip refers to the mean direction that phonons travel after they collide with a boundary and are reflected. Non-zero slip implies that more phonons are reflected forwards than backwards, corresponding to a positive value of the tangential flux at the boundary.  A no-slip condition implies that the same number of phonons are reflected forwards and backwards, leading to the net tangential flux being zero at the boundary.  Slip and no-slip correspond to specular and diffuse modes of scattering, respectively.  Thus, the use of slip conditions enables different scattering modes to be captured in the model.} 

As previously done by other authors \cite{Calvo2018,Zhu2017,Alvarez2009,Sellitto2012,Sellitto2015}, we assume the slip length $\ell^*_s$ is directly proportional to the non-local length $\ell^*$ and write $\ell_s^* = C \ell^*$, where $C$ is a dimensionless parameter that contains information about the interactions between phonons and the boundary of the nanowire. Various forms of $C$ appear in the literature that account for diffusive and specular modes of phonon scattering \cite{Alvarez2009, Zhu2017}, surface roughness \cite{Sellitto2010}, and temperature effects \cite{Sellitto2010, Calvo2018}. Previous slip-based models of heat flow focus on the case of two-dimensional thin films \cite{Sellitto2010,Alvarez2009,Zhu2017} or axisymmetric cylindrical nanowires \cite{Calvo2018}. In both cases, the outer boundary where the slip condition is applied is infinitely long and one dimensional, so the parameter $C$ is independent of space. For the rectangular nanowires considered here, the parameter $C$ is expected to depend on the in-plane transverse coordinates $x^*$ and $y^*$ to account for the behaviour (\emph{e.g.}~increased scattering) of phonons near corners \cite{Ziman2001}. Therefore, we will assume that $C$ is function of the transverse coordinates as well as the temperature so that $C = C(x^*, y^*, T^*)$, where $x^*$ and $y^*$ in this context correspond to points on the boundary. Specific forms of $C$ will be described below.


\subsection{Reduction of the equations}
The governing equations can be simplified by exploiting the long geometry of the nanowire and the similarity between the steady G-K equation \eqref{original_system} and Stokes equations for incompressible viscous flow.  The reduction of the model is analogous to the lubrication approximation in fluid mechanics, which is a common technique for simplifying the governing equations of a viscous fluid confined to a thin-film geometry; see Ockendon and Ockendon \cite{Ockendon1995} for details.  We first introduce dimensionless variables defined as $x=x^*/L^*_1$, $y=y^*/L^*_2$, $z=z^*/L^*_3$, $T=(T^*-T_0^*)/\Delta T^*$, $u=\phi\epsilon^{-1}u^*/w^*_0$, $v=\epsilon^{-1} v^*/w^*_0$, $w=w^*/w^*_0$ and $k(T) = k^*(T^*) / k_0^*$, where $k_0^* = k^*(T_0^*)$ is a reference value of the bulk thermal conductivity and $w_0 = k_0^* \Delta T^* / L_3^*$ corresponds to the classical value of the longitudinal thermal flux.  Using these dimensionless variables in the governing equations shows that the transverse temperature gradients are small, $\pa T / \pa x = O(\phi^{-2} \epsilon^2)$ and $\pa T / \pa y = O(\epsilon^2)$, and may be neglected provided $(\epsilon/\phi)^2 \ll 1$. The classical Fourier flux, in dimensionless form, can then be defined as $w_F = -k \ud T / \ud z$. By writing the dimensionless longitudinal component of the thermal flux in terms of the Fourier flux as $w(x,y,z) = w_F(z) W(x,y)$, it can be shown that $W$ satisfies 
\begin{align}
  W - \Kn^2\left(\phi^2 \padd{W}{x} + \padd{W}{y}\right) = 1,
  \label{app:gk}
\end{align}
where small terms of $O(\epsilon^2)$ and $O(\phi^{-2}\epsilon^2)$ have been neglected.  The parameter $\Kn = \ell^*(T^*) / L_2^*$ is the Knudsen number and determines the dominance of the non-local effects on the heat transport. 
The corresponding boundary conditions are given by
\subeq{
  \label{app:bc}
  \begin{alignat}{2}
    \pad{W}{x} &= 0, &\quad \text{ at } x &= 0, \\
    \pad{W}{y} &= 0, &\quad \text{ at } y &= 0, \\
    W+\phi C\Kn\,\pad{W}{x} &= 0, &\quad \text{ at } x &= 1, \\
    W+ C\Kn\,\pad{W}{y} &= 0, &\quad \text{ at } y &= 1.
\end{alignat}
}
The dimensionless ETC can be defined in terms of the flux $W$ as
\begin{align}
  \frac{k_\text{eff}}{k} = \int_{0}^{1}\int_{0}^{1} W(x,y)\,\ud x \ud y.
  \label{def_keff}
\end{align}
The corresponding dimensional expression $k^*_\eff$ is obtained after multiplying $k_\text{eff}$
with the previously chosen scale $k_0^*$. 

\section{Computation of the effective thermal conductivity}
\label{sec:etc}
When $C$ depends on the transverse coordinates $x$ and $y$, solutions to \eqref{app:gk} and \eqref{app:bc} in general must be obtained using numerical methods. However, in this section, we present analytical expressions for the ETC that are valid under specific limits and assumptions. We first consider the case of a thin nanofilm, characterised by a nanowire with a small cross-sectional aspect ratio $\phi$. We then compute a series expansion of the ETC in the case of spatially uniform slip coefficient $C$. Finally, we examine the small- and large-$\Kn$ limits of the ETC for general slip coefficients.


\subsection{The thin-film limit}

For small cross-sectional aspect ratios, $\phi \ll 1$, the nanowire becomes a thin film. Taking $\phi \to 0$ with $\epsilon \ll \phi$ in \eqref{app:gk} leads to an ordinary differential equation that can be solved using standard methods. The ETC can then be written as
\begin{equation}\label{k_smallphi2}
  \frac{k_\eff}{k}=1 - \int_{0}^{1}\frac{\Kn \tanh(1/\Kn)}{1+C(x,1)\tanh(1/\Kn)}\,\ud x +O(\phi).
\end{equation}
In the case of uniform slip coefficient, the leading-order contributions to \eqref{k_smallphi2} reduce to the expression for the ETC of a thin film obtained by previous authors \cite{Zhu2017,Sellitto2015}. The terms of order $O(\phi)$ account for small changes in the thermal flux due to interactions between phonons and the $x = 1$ boundary. 

For the remainder of this section, we will focus on solutions to the problem in the case when $\phi$ is not small.

\subsection{Spatially uniform slip length}

Under the assumption that the constant $C$ is independent of space, the boundary value problem defined by \eqref{app:gk} and \eqref{app:bc} can be solved analytically by means of an eigenfunction expansion \cite{ebert1965}. The eigenfunctions are of the form $\Psi_{n,m}(x,y)=\cos(\eta_nx)\cos(\mu_my)$, where the eigenvalues $\eta_n$ and $\mu_m$ are respectively the $n^{th}$ and $m^{th}$ solutions of
\begin{equation}\label{eq_mu_eta_rect}
	\cot(\eta_n)=\phi C\Kn\eta_n,\qquad \cot(\mu_m)=C\Kn\mu_m.
\end{equation}
The longitudinal flux can be written as
\begin{align}
  W(x,y) = \sum_{n,m\geq0}\frac{c(\eta_n)c(\mu_m)\cos(\eta_nx)\cos(\mu_my)}{1+\Kn^2(\phi^2\eta_n^2+\mu_m^2)},
  \quad
    c(t)=\frac{2\sin(t)}{t+\sin(t)\cos(t)}.
  \label{eqn:w_C_const}
\end{align}
By inserting \eqref{eqn:w_C_const} into \eqref{def_keff} and integrating term by term, we find that
\begin{align}\label{def_keff_rect}
\frac{k_\eff}{k}=\sum_{n,m\geq0}\frac{c(\eta_n)\sin(\eta_n)c(\mu_m)\sin(\mu_m)}{\eta_n\mu_m(1+\Kn^2(\phi^2\eta_n^2+\mu_m^2))},
\end{align}
which is valid for all values of the Knudsen number $\Kn$. Furthermore, it can be shown that the terms of the series decay as $(nm)^{-2}$ and therefore only a small number of terms is required in practice.


\subsection{Asymptotic approximation for $\Kn \ll 1$}
\label{sec:asy_small_Kn}
Naively taking $\Kn \to 0$ in \eqref{app:gk} shows that $W(x,y) = 1$. However, this solution does not satisfy the slip conditions at $x = 1$ or $y = 1$, which implies there are thin boundary layers near these outer boundaries. In fact, the problem for $\Kn \ll 1$ can be broken down into four spatial regions: (i) the bulk region away from the outer boundaries where $W \simeq 1$; (ii) a boundary layer of width $O(\Kn)$ near $x = 1$; (iii) a second boundary layer of width $O(\Kn)$ near $y = 1$; and (iv) a corner layer near $x = y = 1$ where the two boundary layers overlap. In terms of the integral for the ETC given in \eqref{def_keff}, region (i) gives the classical value of the ETC, since the flux in this region coincides with the classical Fourier flux. Regions (ii) and (iii) give small $O(\Kn)$ corrections that account for phonon collisions with the outer boundaries away from the corners. Region (iv) accounts for even smaller $O(\Kn^2)$ corrections due to corner effects. In the calculations below, we will neglect the corner layer to obtain the first two terms in the small-$\Kn$ expansion of the ETC.

The solution in the boundary layer near $x = 1$ can be obtained by writing $x = 1 - (\phi \Kn) \tilde{x}$ and
$W(x,y) = \tilde{W}(\tilde{x},y)$. Using this scaling in \eqref{app:gk}--\eqref{app:bc} and taking $\Kn \to 0$ gives the problem
\subeq{
  \label{app:bl1}
\begin{align}
  \tilde{W} - \padd{\tilde{W}}{\tilde{x}} = 1,
\end{align}
with boundary conditions
\begin{alignat}{2}
  \tilde{W} - C(1,y) \pad{\tilde{W}}{\tilde{x}} &= 0, &\quad \text{ at } \tilde{x} &= 0, \\
  \tilde{W} &\to 1, &\quad \text{ as } \tilde{x} &\to \infty.
\end{alignat}
}
The solution to \eqref{app:bl1} is given by
\begin{align}
  \tilde{W}(\tilde{x},y) = 1 - \frac{\exp(-\tilde{x})}{1 + C(1,y)}.
\end{align}
A similar approach can be used to determine the solution in the boundary layer at $y = 1$. By writing $y = 1 - \Kn\,\hat{y}$ and $W(x,y) = \hat{W}(x,\hat{y})$ in \eqref{app:gk}--\eqref{app:bc} and taking $\Kn \to 0$, it can be shown that $\hat{W}$ is given by
\begin{align}
  \hat{W}(x,\hat{y}) = 1 - \frac{\exp(-\hat{y})}{1 + C(x,1)}.
\end{align}
The composite asymptotic solution, which is a single expression that includes contributions from each of the spatial regions (minus those from the corner layer), can be written as
\begin{align}
  W(x,y) = 1 - \frac{\exp[-(1 - x) / (\phi \Kn)]}{1 + C(1,y)} - \frac{\exp[-(1 - y)/\Kn]}{1 + C(x,1)}.
  \label{app:W_comp}
\end{align}
Using \eqref{app:W_comp} in \eqref{def_keff} shows that the ETC is given by
\begin{align}
  \frac{k_\text{eff}}{k} = 1 - \Kn \left[\phi \int_{0}^{1}\frac{\ud y}{1 + C(1,y)} +\int_{0}^{1}\frac{\ud x}{1 + C(x,1)}\right] + O(\Kn^2).
  \label{keff_smallKn}
\end{align}
If the slip length is spatially uniform, Eqn.~\eqref{keff_smallKn} reduces to $k_\text{eff} / k = 1 - \Kn (1 + \phi) / (1 + C) + O(\Kn^{2})$.

\subsection{Asymptotic approximation for $\Kn \gg 1$}
\label{sec:asy_large_Kn}
The strong influence of boundary effects when the Knudsen number is large leads to a substantial reduction in the
thermal flux. As discussed in Calvo \emph{et al.}~\cite{Calvo2018}, the flux scales like $W = O(\Kn^{-1})$ in the limit as $\Kn \to \infty$. Thus, after writing $W(x,y) = \Kn^{-1} \bar{W}(x,y)$ in \eqref{app:gk}--\eqref{app:bc}, the problem for the rescaled flux $\bar{W}$ is given by
\begin{align}
  \Kn^{-1} \bar{W} - \Kn\left(\phi^2 \padd{\bar{W}}{x} + \padd{\bar{W}}{y}\right) = 1,
  \label{app:gk2}
\end{align}
subject to
\subeq{
  \label{app:bc2}
  \begin{alignat}{2}
    \pad{\bar{W}}{x} &= 0, &\quad \text{ at } x &= 0, \\
    \pad{\bar{W}}{y} &= 0, &\quad \text{ at } y &= 0, \\
    \bar{W}+\phi C\Kn\,\pad{\bar{W}}{x} &= 0, &\quad \text{ at } x &= 1, \\
    \bar{W}+ C\Kn\,\pad{\bar{W}}{y} &= 0, &\quad \text{ at } y &= 1.
\end{alignat}
}
By integrating the G-K equation \eqref{app:gk2} over the cross-sectional area and using the boundary conditions
\eqref{app:bc2}, we find that
\begin{align}
  \Kn^{-1} \int_{0}^{1}\int_{0}^{1} \bar{W}(x,y)\,\ud x \ud y +
  \left[\phi \int_{0}^{1}C^{-1}(1,y)\bar{W}(1,y)\,\ud y + \int_{0}^{1} C^{-1}(x,1) \bar{W}(x,1)\,\ud x\right]
  = 1.
  \label{app:int}
\end{align}
Provided that $C(x,y) \gg \Kn^{-1}$, taking $\Kn \to \infty$ in \eqref{app:gk2}--\eqref{app:bc2} shows that $\bar{W}$ is constant in space. The integral relation in \eqref{app:int} then enables $\bar{W}$ and hence $W$ to be easily determined.
Therefore, we have that
\begin{align}
  \frac{k_\text{eff}}{k} = \Kn^{-1}\left[\phi \int_{0}^{1}C^{-1}(1,y)\,\ud y + \int_{0}^{1} C^{-1}(x,1)\,\ud x\right]^{-1}
  + O(\Kn^{-2})
  \label{keff_largeKn}
\end{align}
as $\Kn \to \infty$.
In the case of spatially uniform slip lengths, Eqn.~\eqref{keff_largeKn} reduces to $k_\text{eff} / k = C \Kn^{-1} / (1 + \phi) + O(\Kn^{-2})$.


\section{Results and discussion}
\label{sec:res}
\begin{figure}
  \centering
  \subfigure[$w/w_F$ for $\Kn = 0.1$]{\includegraphics[width=0.32\textwidth]{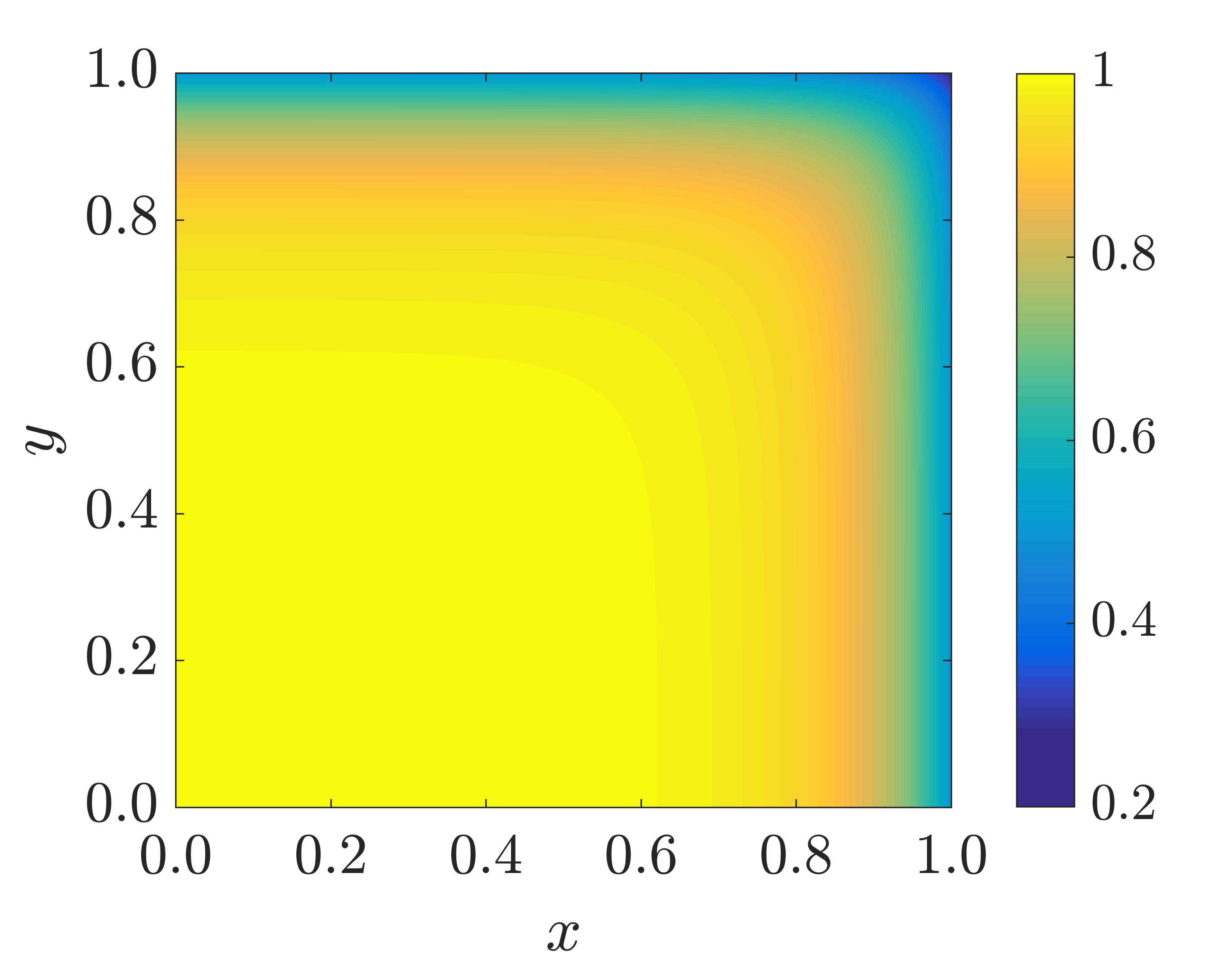}}
  \subfigure[$w / w_F$ for $\Kn = 10$]{\includegraphics[width=0.32\textwidth]{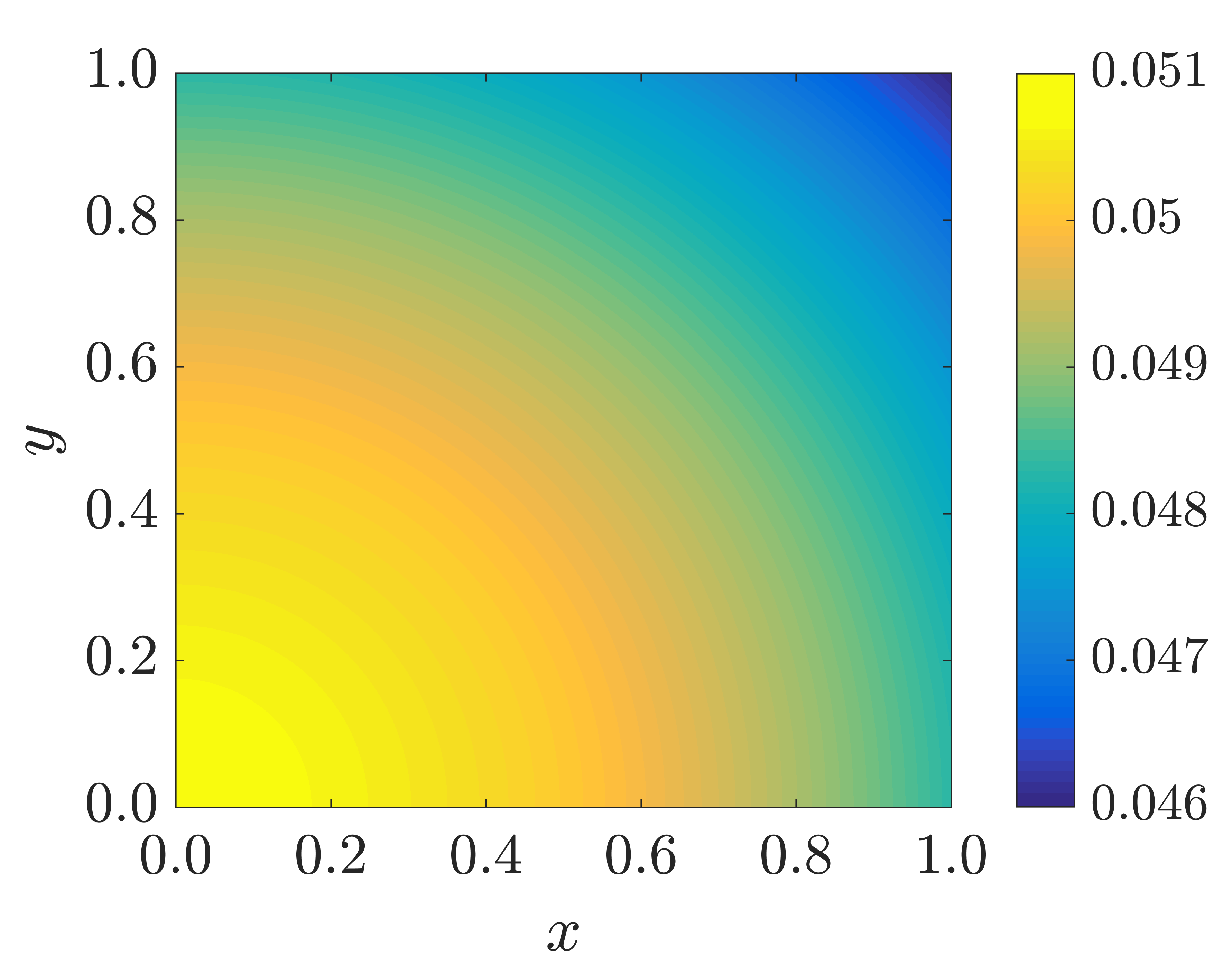}}
  \subfigure[$k_\text{eff} / k$]{\includegraphics[width=0.32\textwidth]{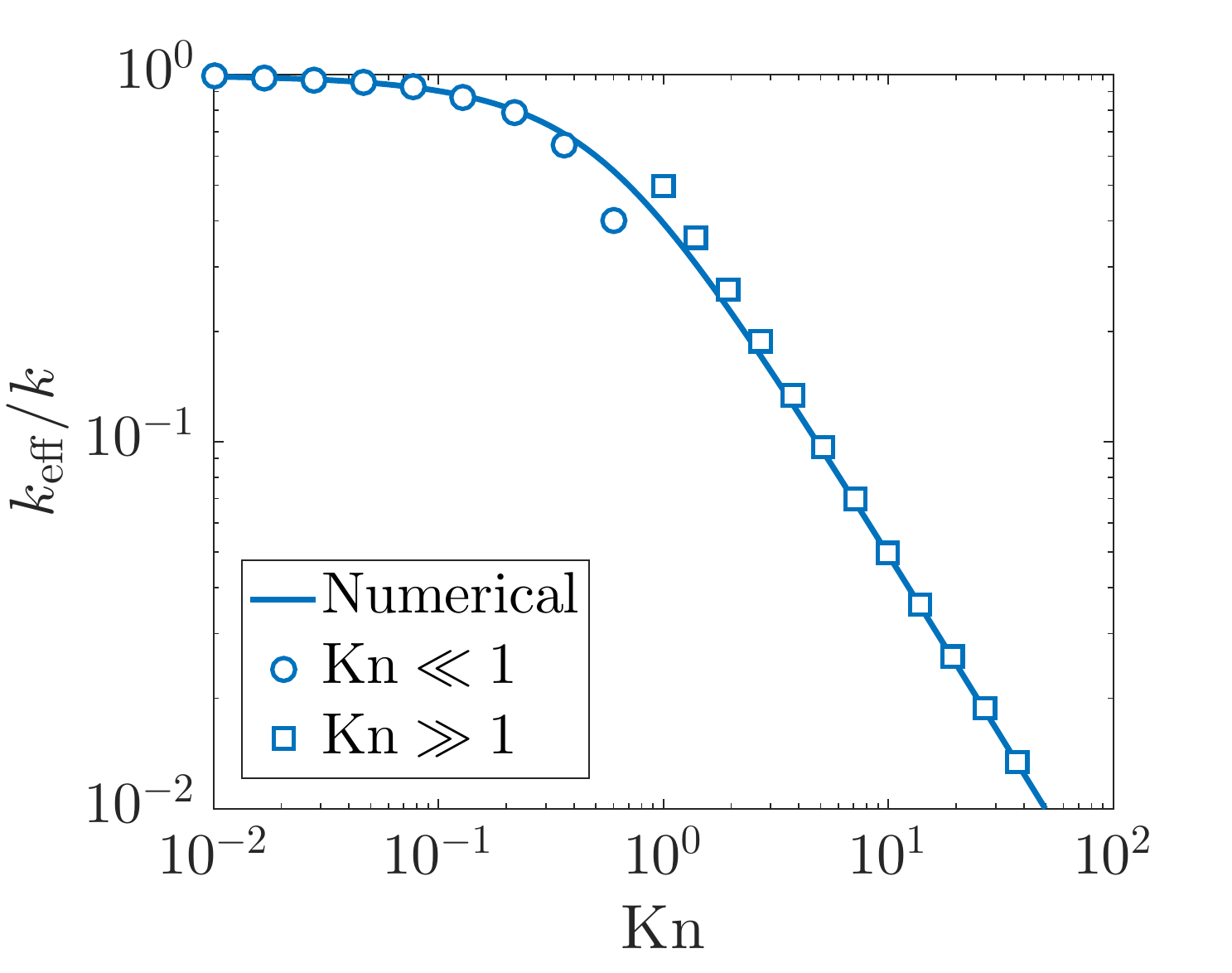}}
  \caption{(a)--(b): Contour maps of the normalised longitudinal thermal flux $w / w_F$, where $w_F = -k\, \ud T / \ud x$ is the classical Fourier flux, for two values of the Knudsen number: $\Kn = 0.1$ and $\Kn = 10$. (c):
    The dependence of the normalised ETC $k_\text{eff} / k$ on the Knudsen number. The asymptotic solution for
    $\Kn \ll 1$ and $\Kn \gg 1$ is given by \eqref{keff_smallKn} and \eqref{keff_largeKn}, respectively.
    In all panels, $\phi = 1$ and $C = 1$.}
  \label{fig:wk}
\end{figure}

We first examine how changes to the Knudsen number affect both the longitudinal thermal flux and the ETC for fixed values of $\phi = 1$, corresponding to square nanowires, and $C = 1$. Panels (a) and (b) of Figure \ref{fig:wk} show contour maps of the normalised flux $w / w_F$
for two values of the Knudsen number: $\Kn = 0.1$ (Fig.~\ref{fig:wk}~(a)) and $\Kn = 10$ (Fig.~\ref{fig:wk}~(b)). The corresponding normalised ETC, $k_\text{eff} / k$, is shown as a function of the Knudsen number in Fig.~\ref{fig:wk} (c).

For small Knudsen numbers, the non-local length is much less than the cross-sectional dimensions of the nanowire. Phonons in the bulk will frequently collide with each other, resulting in diffusive transport of heat that is not strongly influenced by the finite geometry of the cross section. Indeed, as shown in Fig.~\ref{fig:wk} (a), the normalised flux in the bulk is close to unity and thus coincides with the diffusive Fourier flux. However, there is a region of ballistic transport near the exterior boundaries of the nanowire, where phonon-boundary scattering becomes relevant. In this ballistic region, there is a rapid decrease in the flux from its bulk value of approximately $w_F$ across dimensional length scales of order $\ell^*$ (or non-dimensional length scales of order $\Kn$). Surprisingly, the thermal flux does not tend to zero at the exterior boundaries, despite the fact that naively taking $\Kn \to 0$ in the boundary conditions \eqref{app:bc} appears to yield the no-slip condition. However, the gradient of the flux scales like $\Kn^{-1}$ near the exterior boundaries, cancelling the $\Kn$ in the slip coefficient and preventing the no-slip condition from being recovered in the small-$\Kn$ limit. This is shown explicitly in Sec.~\ref{sec:asy_small_Kn}.

Given that the non-classical, ballistic region of transport is confined to a thin boundary layer when the Knudsen number is small, the ETC is well approximated by its classical value, $k_\text{eff} / k \simeq 1$, for $\Kn \ll 1$. In fact, as shown in Sec.~\ref{sec:asy_small_Kn}, the small $O(\Kn)$ correction in the asymptotic approximation \eqref{keff_smallKn} accounts for the reduced heat transport that occurs in the boundary layers. The asymptotic approximation \eqref{keff_smallKn} is plotted as circles in Fig.~\ref{fig:wk} (c) and is in excellent agreement with the numerically computed curve (solid line) obtain from the full model.

For large Knudsen numbers, the non-local length greatly exceeds the dimensions of the cross section. In this case, phonons frequently collide with the exterior boundary, leading to a substantial reduction in the transport of thermal energy across the nanowire; see Fig.~\ref{fig:wk} (b). As discussed by Calvo \emph{et al}.~\cite{Calvo2018} in the case of circular nanowires, the flux for large Knudsen numbers is of order $\Kn^{-1}$. Furthermore, due to the large effective slip coefficient, the transverse gradients in the flux are small, resulting in a roughly uniform thermal flux over the cross-sectional area. The greatly reduced thermal transport in the large-$\Kn$ regime is reflected by the relatively low values of the ETC, which is also proportional to $\Kn^{-1}$, as seen from the asymptotic result \eqref{keff_largeKn}. As noted by previous authors \cite{Alvarez2009, Calvo2018}, the slip boundary condition is crucial for obtaining an ETC that scales with $\Kn^{-1}$ for $\Kn \gg 1$ from the hydrodynamic model, which is required for agreement with experimental data (see, for example, Ref.~\cite{berman1955} or the data below). The asymptotic expression for the ETC \eqref{keff_largeKn} is shown as the squares in Fig.~\ref{fig:wk} (c) and is in excellent agreement with the numerical calculations. 

Figures \ref{fig:wk} (a) and (b) show that, regardless of the value of the Knudsen number, the thermal flux attains its minimum value at the corner of the nanowire. This minimum is a consequence of the imperfect slip that occurs at the boundaries, which leads to greater resistance to heat flow (\emph{i.e.} more phonon reflections) in the corners. 

\begin{figure}
  \centering
  \subfigure[]{\includegraphics[width=0.49\textwidth]{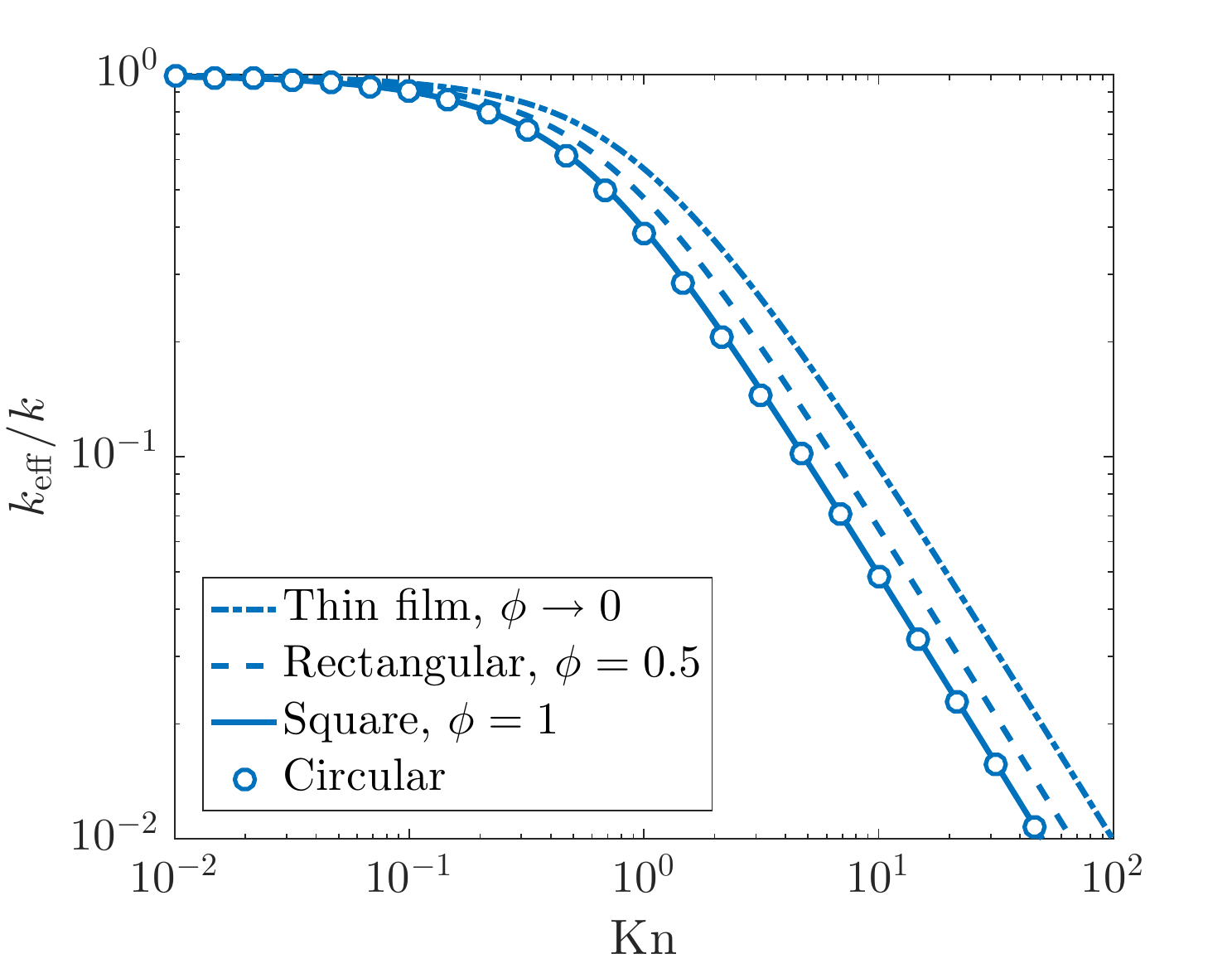}}
  \subfigure[]{\includegraphics[width=0.49\textwidth]{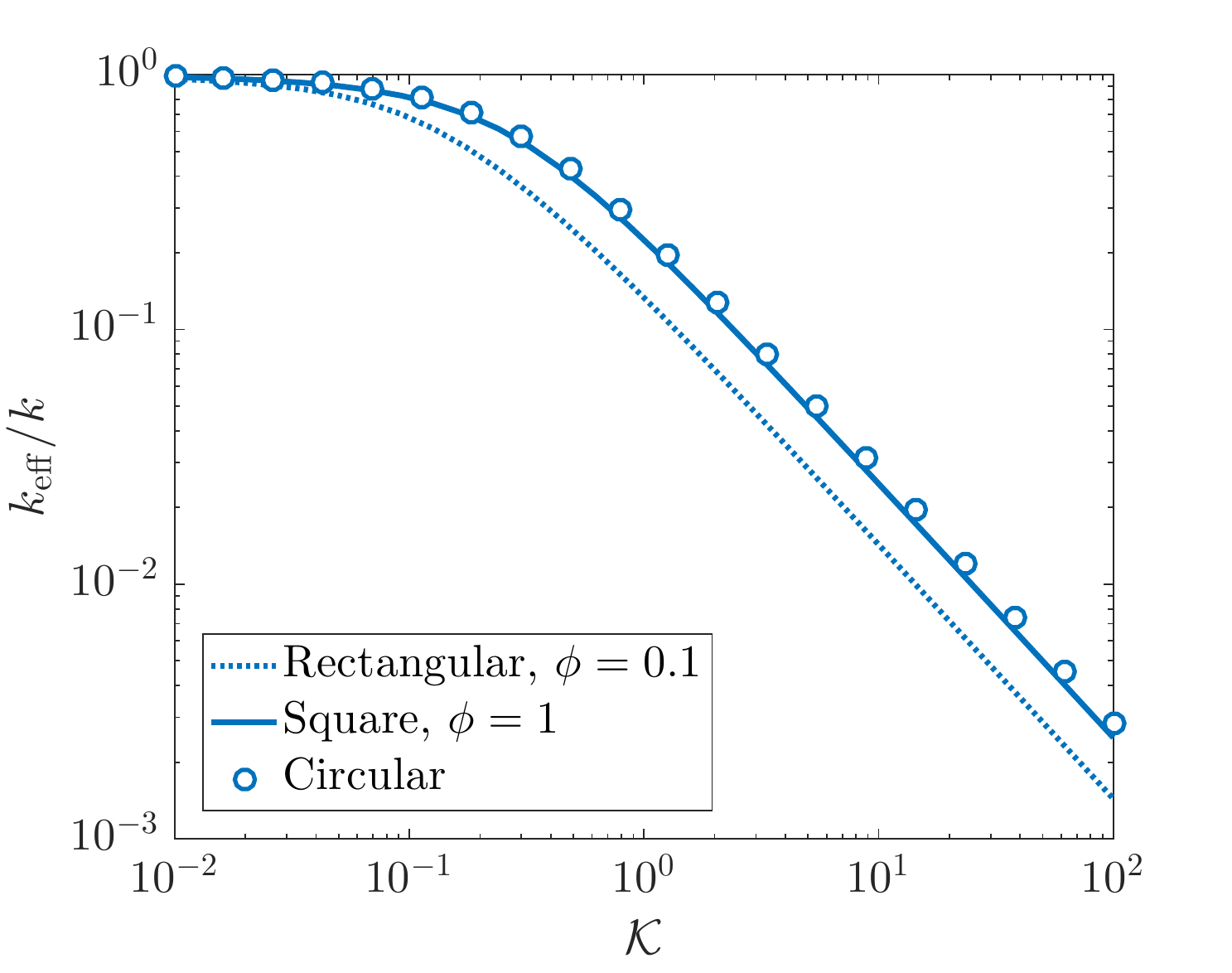}}
  \caption{The dependence of the normalised ETC $k_\text{eff} / k$ on (a) the Knudsen number $\Kn$ and (b) the
    effective Knudsen number $\mathcal{K} = \ell^* / \sqrt{A^*}$ for various cross-sectional aspect ratios
    $\phi$. Also shown in the normalised ETC of a circular nanowire \cite{Calvo2018}. Fixed values of $\mathcal{K}$
    correspond to fixed values of the cross-sectional area $A^*$. In both panels a value of $C = 1$ was used.}
  \label{fig:k_phi}
\end{figure}

We now examine how changes to the cross-sectional geometry affect the ETC for a fixed value of $C = 1$. Figure \ref{fig:k_phi} (a) shows the ETC of a two-dimensional thin film ($\phi \to 0$), a rectangular nanowire ($\phi = 1/2$), and a square nanowire ($\phi = 1$). For comparison purposes, we also plot the ETC for a circular nanowire obtained by Calvo \emph{et al}.~\cite{Calvo2018}. As the cross-sectional aspect ratio $\phi$ decreases from one to zero, the ETC increases for all values of the Knudsen number. To understand the reason for this increase, it is helpful to recall that $\phi$ is defined in terms of the dimensions of the cross section as $\phi = L_2^* / L_1^*$ and the Knudsen number is defined as $\Kn = \ell^* / L_2^*$. Thus, decreasing $\phi$ for fixed $\Kn$ is equivalent to increasing $L_1^*$ for fixed $L_2^*$. The consequential increase in cross-sectional area essentially leads to the corners of the nanowire becoming further separated, thereby reducing their negative impact on the thermal flux. In the case of a thin film ($\phi \to 0$), there are no corners to reduce the flux; hence, the ETC of a thin film is greater than the ETC of square and rectangular nanowires. Interestingly, however, the ETC of a circular nanowire, which has no corners at all, is virtually identical to the ETC of a square nanowire. 
\rev{This similarity can be rationalised in terms of the hydraulic diameter of the nanowires, defined as $D^* = 4 A^* / P^*$, where $A^*$ and $P^*$ are the (dimensional) cross-sectional area and perimeter, respectively.  For the rectangular nanowires considered here, we can write $D_\text{rect}^* = 4 L_2^* / (1 + \phi)$; for circular nanowires of radius $R^*$, we find that $D_\text{circ}^* = 2 R^*$. The hydraulic diameters of square and circular nanowires will be equal if the half-height $L_2^*$ is the same as the radius $R^*$. It is also possible to introduce an equivalent Knudsen number based on the hydraulic diameter, $\Kn' = 2 \ell^* / D^*$.  Unlike the Knudsen number $\Kn = \ell^* / L_2^*$, the equivalent Knudsen number $\Kn'$ simultaneously captures the $x$ and $y$ dimensions of the nanowire. For square and circular nanowires, the equivalent Knudsen numbers $\Kn'$ are the same, and both are equal to the original Knudsen number $\Kn$. Furthermore, by extending the asymptotic analysis of Sec.~\ref{sec:asy_large_Kn} to nanowires of arbitrary cross section, we find that the ETC, in the limit of large equivalent Knudsen number and constant slip coefficient $C$, can be written as 
\begin{align}
k_\text{eff} / k \sim (C / 4) (D^* / \ell^*),
\label{eqn:k_eff_D}
\end{align}
revealing that the limiting behaviour of the ETC is indeed the same for square and circular nanowires.  Equation \eqref{eqn:k_eff_D} shows that, regardless of the nanowire geometry, the ETC for large Knudsen numbers is set by the non-local length and the hydraulic diameter.}

To better understand the role of the nanowire geometry, we now investigate how the ETC varies with $\phi$ for fixed cross-sectional areas $A^* = 4 L_1^* L_2^*$, keeping $C = 1$. In order to maintain a constant area $A^*$ in the dimensionless model, we need to account for how the Knudsen number $\Kn = \ell^* / L_2^*$ varies with $\phi$ through $L_2^*$. To do so, we first note that $L_2^*$ can be written as $L_2^* = \sqrt{\phi A^* / 4}$. The Knudsen number can then be written in terms of the aspect ratio $\phi$ as $\Kn = (2 / \sqrt{\phi}) \mathcal{K}$, where $\mathcal{K} = \ell^* / \sqrt{A^*}$ is an effective Knudsen number that is independent of the shape of the cross section and which remains constant as $\phi$ varies. \rev{Similar to the equivalent Knudsen number $\Kn'$, the effective Knudsen number $\mathcal{K}$ also accounts for the full two-dimensional geometry of the nanowire rather than the length along one particular dimension.}  Figure \ref{fig:k_phi} (b) compares the ETC of a rectangular and a square nanowire with $\phi = 0.1$ and $\phi = 1$, respectively, over a range of $\mathcal{K}$. Contrary to the results of Fig.~\ref{fig:k_phi} (a), here we see that for a fixed value of $\mathcal{K}$, corresponding to a fixed area $A^*$, increases in the cross-sectional aspect ratio $\phi$ lead to increases in the ETC. This relationship stems from the fact that as the aspect ratio $\phi$ decreases, the exterior boundaries at $y^* = \pm L_2^*$ become closer together and the bulk heat flow becomes dominated by phonon scattering at these surfaces. The ETC of the circular nanowire is also shown in Fig.~\ref{fig:k_phi} (b) and is found to be slightly greater than the square for all values of $\mathcal{K}$ due to the lack of corners. In plotting the ETC for the circular nanowire, the Knudsen number has been written as $\Kn = \sqrt{\pi}\,\mathcal{K}$. Thus, for a fixed cross-sectional area, the circular nanowire is the most efficient transporter of thermal energy, with the thin film being the worst.

\begin{figure}
  \centering
  \includegraphics[width=0.49\textwidth]{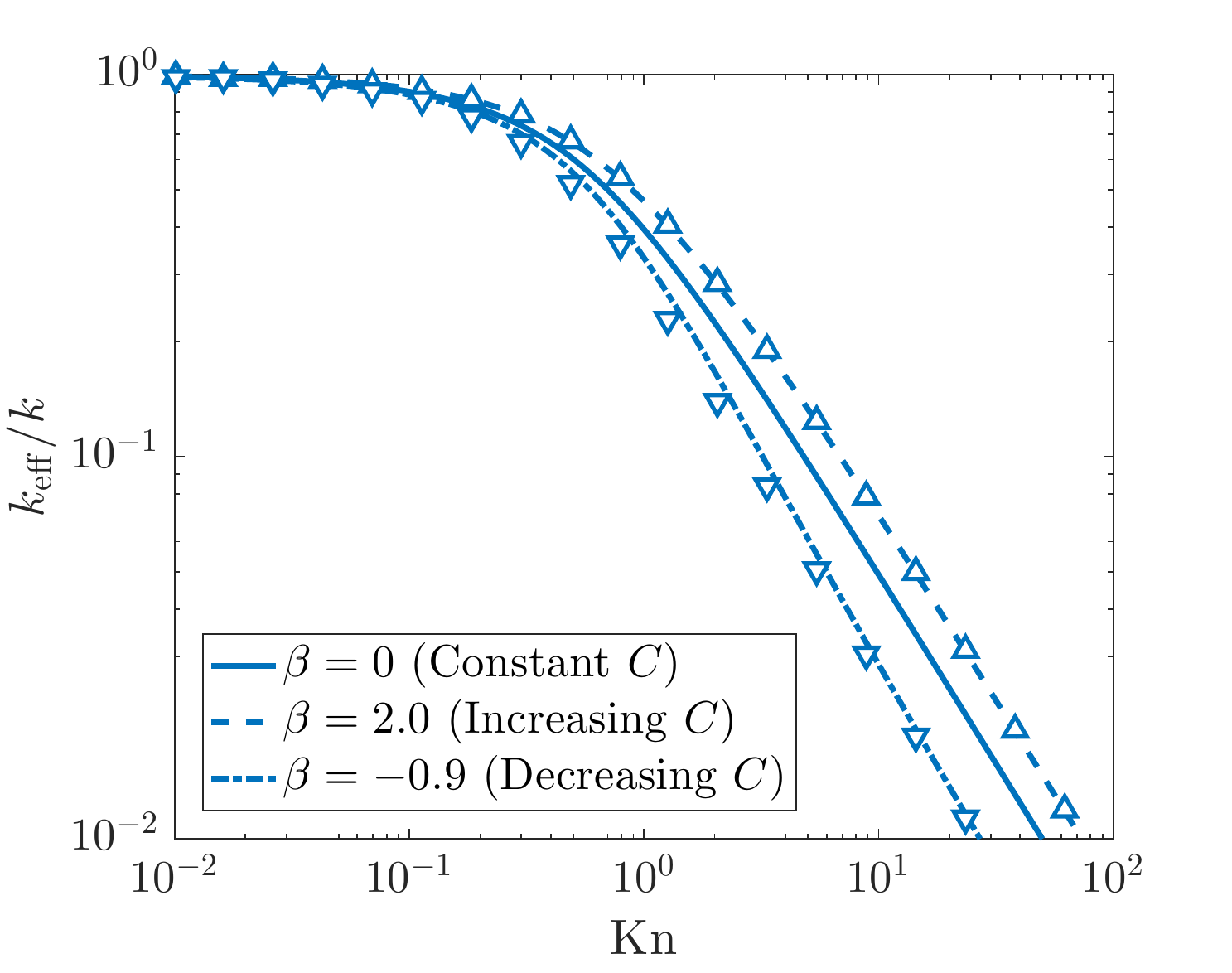}
  \caption{The influence of non-uniform slip coefficients $C$ on the normalised ETC $k_\text{eff} / k$ of
    a square nanowire ($\phi = 1$). Lines
    correspond to the ETC computed using the slip coefficient in \eqref{eqn:non_uni_C}. The cases $\beta = 0$,
    $\beta = 2$, and $\beta = -0.9$ correspond slip coefficients that are constant, increasing towards a corner,
    and decreasing towards a corner. The upwards- and downwards-pointing triangles denote the ETC computed using
    the equivalent slip coefficients \eqref{eqn:C_e_pos} and \eqref{eqn:C_e_neg}, respectively.}
  \label{fig:k_C}
\end{figure}

The influence of the slip coefficient $C$ on thermal transport has been discussed by Zhu \emph{et al}.~\cite{Zhu2017} in the context of a thin film ($\phi \to 0$) under the assumption that $C$ is a constant. In summary, larger values of $C$ lead to less resistance to heat flow and thus to greater values of the thermal flux and ETC. A distinguishing feature of rectangular nanowires is that the slip coefficient $C$ may depend on the boundary coordinates to account for the change in phonon behaviour near corners. To explore this behaviour with our model, we consider a square nanowire ($\phi = 1$) and a simple form of the slip coefficient given by
\begin{align}
  C(x,y) = \alpha + \beta(x^2 + y^2 - 1),
  \label{eqn:non_uni_C}
\end{align}
where $\alpha$ and $\beta$ are constants that satisfy $\alpha > 0$ and, to ensure $C$ remains positive, $\beta > -\alpha$. The parameter $\beta$ controls the curvature of the slip coefficient. If $\beta < 0$ ($\beta > 0$) then $C$ decreases (increases) towards a corner. The lines in Fig.~\ref{fig:k_C} represent numerical calculations of the ETC using finite differences in the case of $\alpha = 1.0$ and $\beta = -0.9$, $0$, and $2.0$. Slip coefficients that decrease (increase) towards a corner lead to a corresponding decrease (increase) in the ETC, as these amplify (counteract) the resistance to thermal transport occurring at the corners. Qualitatively, the dependence of the ETC on the Knudsen number is similar to the case when the slip coefficient is constant. This naturally leads to the question of whether it is possible to define an equivalent, \emph{constant} slip coefficient, $C_\text{equiv}$, that yields quantitatively similar results to the non-uniform slip coefficient \eqref{eqn:non_uni_C}. One possibility for obtaining an equivalent slip coefficient is to match the limiting behaviour of the ETC for large Knudsen numbers. From Sec.~\ref{sec:asy_large_Kn}, the large-$\Kn$ behaviour of the ETC for a constant slip coefficient in the case of a square nanowire is given by $k_\text{eff} / k \sim (C_\text{equiv}/2) \Kn^{-1}$. By inserting the slip coefficient \eqref{eqn:non_uni_C} into \eqref{keff_largeKn}, performing the integration, and equating the result to $(C_\text{equiv}/2)\Kn^{-1}$, we find an equivalent slip constant given by
\subeq{
\begin{alignat}{2}
  C_\text{equiv} &= \frac{\sqrt{\alpha \beta}}{{\rm{arctan}}\left(\sqrt{\beta/\alpha}\right)}, &\quad &\beta > 0,
  \label{eqn:C_e_pos}
  \\
  C_\text{equiv} &= \frac{\sqrt{\alpha |\beta|}}{{\rm{arctanh}}\left(\sqrt{|\beta|/\alpha}\right)}, &\quad &\beta < 0.
  \label{eqn:C_e_neg}
\end{alignat}
}
The upwards- and downwards-pointing triangles in Fig.~\ref{fig:k_C} show the ETC computed with the equivalent slip coefficients given by \eqref{eqn:C_e_pos} and \eqref{eqn:C_e_neg}, respectively. The quantitative agreement between the ETCs computed using the equivalent and non-uniform slip conditions is remarkable. For $\alpha = 1$, we find that $C_\text{equiv} \simeq 1.48$ when $\beta = 2$ and $C_\text{equiv} \simeq 0.52$ when $\beta = -0.9$. Thus, local increases (decreases) in the slip coefficient due to corner effects are equivalent to increases (decreases) in the bulk slip coefficient, highlighting the non-local nature of nanoscale heat transport.

To validate the model, we compare theoretical predictions of the ETC against experimental measurements carried out using silicon wires by Inyushkin \emph{et al}.~\cite{Inyushkin2004}. The rectangular wires are of dimension $2L_1^* = 3.12$ mm, $2L_2^* = 2.00$ mm, and $L_3^* = 20.5$ mm. Although the wires are macroscopic in size, they are held at temperatures for which their dimensions are much smaller than the MFP of phonons. The longitudinal and cross-sectional aspect ratios of the wire are $\epsilon \simeq 0.098$ and $\phi \simeq 0.64$, respectively. Temperature gradients ranging from $0.01$ K to $0.2$ K were imposed across the length of the wire. The temperature at the end point of the wire was held between 0.5 K and 293 K, allowing both the \rev{hydrodynamic} (low temperature) and diffusive (high temperature) regimes of thermal transport to be explored. Numerically evaluating the theoretical ETC requires knowledge of the temperature-dependent non-local length and the bulk thermal conductivity. These quantities have been extracted from first-principle calculations based on the KCM framework \cite{Torres2017b} using open-source code \cite{Torres2017a}, as described in \ref{app:kcm}.

\begin{figure}
  \centering
  \subfigure[]{\includegraphics[width=0.49\textwidth]{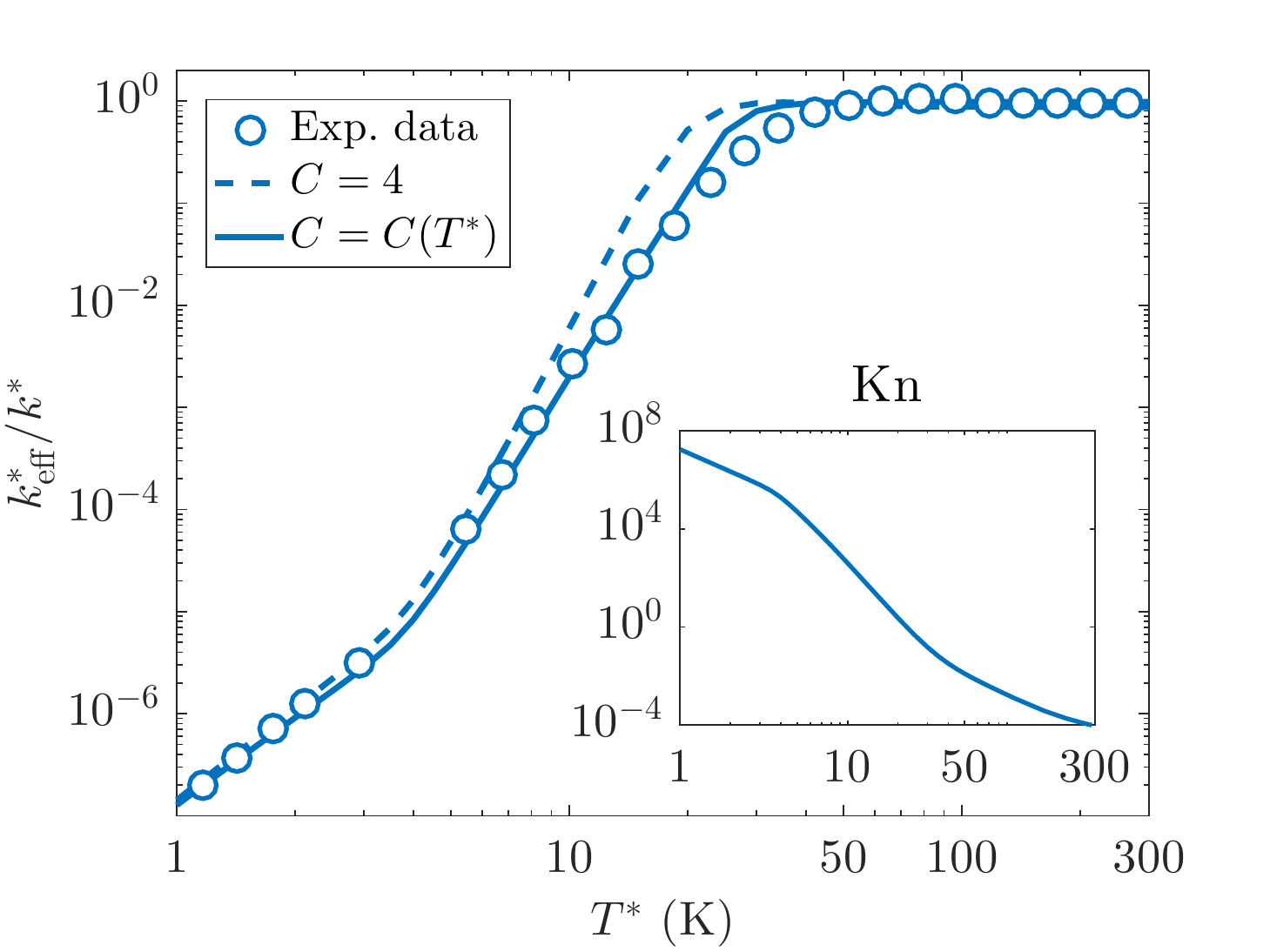}}
  \subfigure[]{\includegraphics[width=0.49\textwidth]{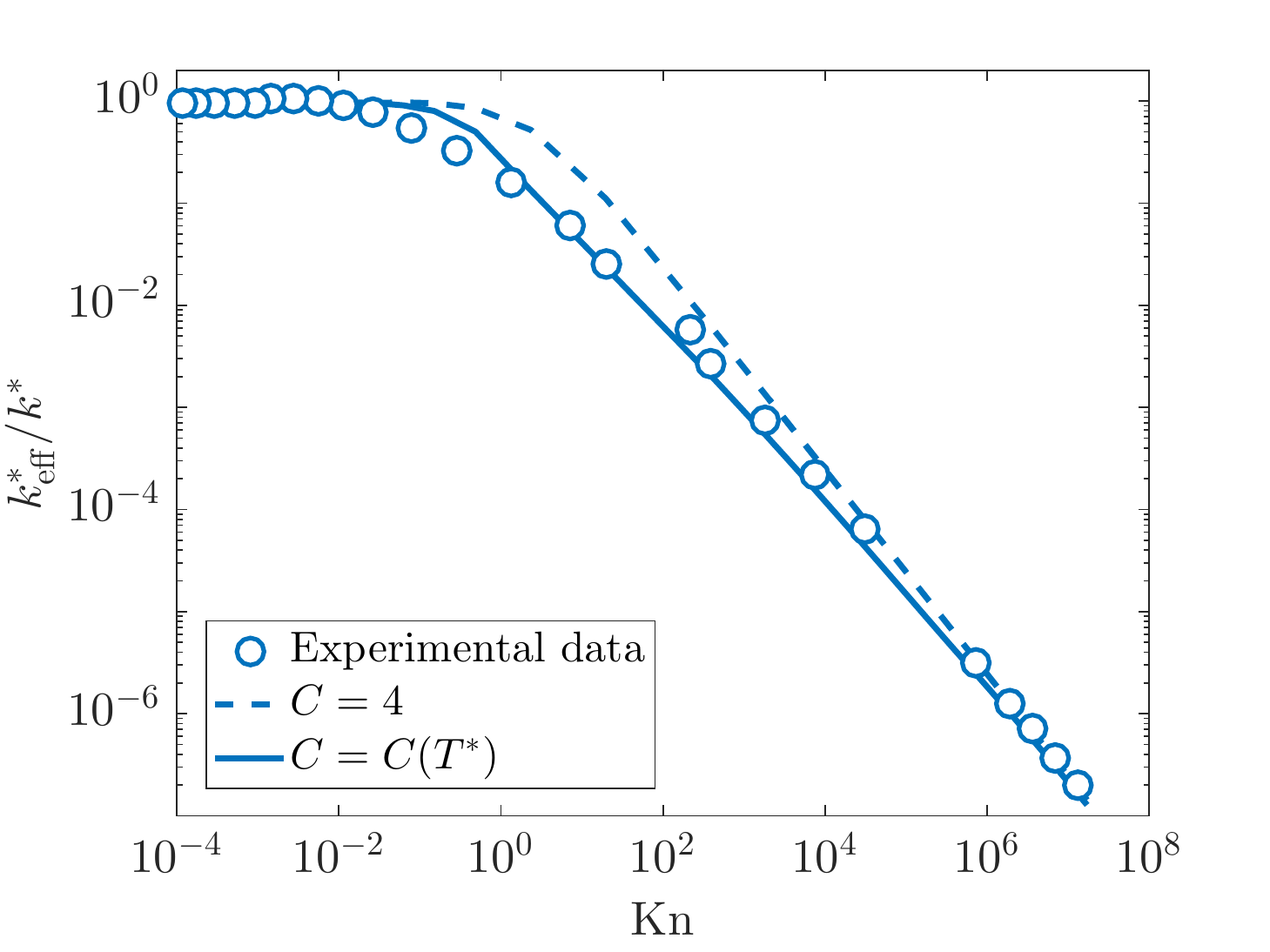}}
  \caption{Comparison experimental measurements (circles) of the ETC for rectangular wires
    \cite{Inyushkin2004} with theoretical predictions computed using a constant
    $C = 4$ (dashed lines) and a temperature-dependent $C(T^*) = 4\,\exp(-T^* / T_c^*)$ (solid lines) slip
    coefficients. The inset of panel (a) shows the temperature dependence of the Knudsen number.}
  \label{fig:exp}
\end{figure}

The experimental measurements of the ETC, normalised by the bulk thermal conductivity, are shown as circles in Fig.~\ref{fig:exp}. Panels (a) and (b) of this figure plot the ETC as a function of the temperature and Knudsen number, respectively. The inset shows the temperature dependence of the Knudsen number, allowing Figs.~\ref{fig:exp} (a) and (b) to be connected. The Knudsen number is small for temperatures exceeding 50 K ($\Kn \simeq 0.013$). Consequently, the experimental measurements of the ETC coincide with the bulk thermal conductivity. However, as the temperature is decreased below 50 K, the Knudsen number becomes order one in magnitude and the ETC begins to rapidly decrease from its bulk value. The large-$\Kn$ regime is entered as the temperature falls below 10 K ($\Kn \simeq 400$)
and the measurements of the ETC follow the $\Kn^{-1}$ scaling that was derived from the $\Kn \gg 1$ limit of the model.

We first attempt to capture the experimental data using the model with a constant slip coefficient. Although this is the simplest choice available, it is also motivated by the discussion surrounding Fig.~\ref{fig:k_C}, which argues that spatially non-uniform slip coefficients can yield quantitatively similar results. An estimate of the slip coefficient can be obtained by fitting the experimental data points for $\Kn > 10^6$ to the large-$\Kn$ limit of the ETC given by \eqref{keff_largeKn}. This procedure produces the value of $C \simeq 4$. This value of $C$ agrees with the results obtained using the Mathiessen rule in previous works \cite{Torres2017b,Zhang2007}: in the limit $\Kn\to\infty$ the effective thermal conductivity for a thin film $(\phi\to0)$ is $k^*_\eff\approx(2H^*/\ell^*)k^*$, where $H^*=2L_{2}^*$ is the thickness of the film. The dashed lines in the main panels of Fig.~\ref{fig:exp} denote the theoretical prediction of the ETC using $C = 4$. The model prediction is remarkably accurate for Knudsen numbers above $10^2$ ($T^* < 10$ K). The model also gives accurate predictions for Knudsen numbers below $10^{-2}$ ($T^* > 50$ K), where the bulk thermal conductivity is recovered and the value of the slip coefficient becomes irrelevant. The over-estimation of the ETC by the model across intermediate values of the Knudsen number ranging from $10^{-2}$ to $10^{2}$ (temperatures between 10 K and 50 K) is attributed to the assumption of a constant slip coefficient, which is too large in this region.  \rev{Evidently, the slip coefficient must begin to decrease as the temperature increases.  The temperature dependence of the slip coefficient can be rationalised in terms of the dynamics of phonon-boundary scattering.  As the temperature increases, diffuse scattering becomes increasingly dominant over specular scattering \cite{Sellitto2010}, with specular scattering becoming negligible at a critical temperature $T_c^*$ depending on the roughness of the nanowire and phonon wavelength \cite{Asheghi1998}.  Diffuse scattering emits phonons in and against the direction of the temperature gradient and therefore leads to a vanishing thermal flux at the exterior boundary of the wire, which is captured by a slip coefficient that tends towards zero for temperatures $T^*$ exceeding the critical value $T_c^*$.  Therefore, we propose a simple exponential form for the slip coefficient given by $C(T^*) = 4\,\exp(-T^* / T_c^*)$. We treat $T_c$ as a fitting parameter whose value will depend on the physical and geometrical (roughness) characteristics of the nanowire.}  From a visual inspection of Fig.~\ref{fig:exp} (a), noticeable differences between the experimental and theoretical predictions with $C = 4$ begin to occur when $T^* \simeq 9$ K, marking the temperature at which the slip coefficient begins to decrease. Taking $T_c^* = 9$ in the exponential form of $C$ yields the ETC predictions given by the solid lines in Fig.~\ref{fig:exp}, which show a considerable improvement over the estimates with $C = 4$. The model gives accurate predictions over nearly the entire temperature range; however, some errors do persist for temperatures around 30 K. Refined estimates of the ETC can likely be obtained using a physical model for the slip coefficient that more accurately describes the nature of phonon-boundary interactions. For now, we leave formulating such a model as a key area of future research.

The existing discrepancies of our model with respect to experimental data can also be understood from the Truncated L\'evy Flights (TLF) framework \cite{Vermeersch2015}. According to the TLF theory, when the dominant characteristic length of the thermal transport is reduced (\emph{i.e.}, the phonon MFP), phonon transport moves from the diffusive to ballistic (or hydrodynamic) regime through a superdiffusive transport regime, which has an ETC that is smaller than the expected from the simple diffusive-ballistic transition. Notice that here the analogy ballistic-hydrodynamic refers to a transport regime dominated by a single characteristic length scale.

In silicon, when the temperature decreases below 50~K, phonons with a MFP larger than $\sim$3~mm (of the order of width and height of the slab) begin to appear. While at high temperature (diffusive regime) those phonons do not have a significant contribution, at very low temperatures (ballistic or hydrodynamic regime) they are removed from the model by the boundary conditions. In the diffusive-ballistic transition around 7--50~K, where these phonons can be important, the present model over-estimates the experimental data, as a single characteristic non-local length in the transport equation is not able to capture the superdiffusive transition.

\section{Conclusion}
\label{sec:conc}
We use a hydrodynamic model of thermal transport, based on the Guyer--Krumhansl equation, with a slip boundary condition to calculate the ETC of rectangular nanowires. An analytical solution for the ETC is obtained in the case of a constant slip coefficient by writing the thermal flux as an eigenfunction expansion. For spatially non-uniform slip coefficients, asymptotic methods are employed to calculate the leading approximations valid for small and large Knudsen numbers. An attractive feature of the asymptotic approach is that it can be straightforwardly applied to nanowires with arbitrary cross section, enabling the ETC to be calculated with ease for a range of geometries that might be non-trivial to implement using standard computational techniques.

A key advantage of the proposed model is that it is based on a minimal number of free parameters. Both the non-local length and the bulk thermal conductivity can be calculated as a function of temperature using open-source software \cite{Torres2017a} and considered as known quantities. For sufficiently small temperatures, the correct slip coefficient is approximately four, in accordance with previous work \cite{Torres2017b,Zhang2007}. This enables the ETC predictions to be obtained without any fitting parameters in this regime. To account for changes in the slip coefficient with increasing temperature, a simple exponential form with a single fitting parameter is used. A comparison between experiment and theory reveals that the model is able to accurately predict the ETC across a wide range of temperatures, which is particularly remarkable given the simple nature of the underlying model. Future studies will focus on deriving an expression for the slip coefficient using a detailed physical model. By writing the slip coefficient in terms of known physical quantities rather than fitting parameters, the predictability of the proposed hydrodynamic model can be greatly improved, yielding accurate estimates for the ETC across arbitrary temperature ranges and providing key insights into how thermal energy will be distributed across nanosystems.

\section*{Acknowledgements}

M.~C.~acknowledges that the research leading to these results has received funding from ‘la Caixa’ Foundation. M.~H. acknowledges funding from the European Union's Horizon 2020 research and innovation programme under grant agreement No. 707658. P.~T. and F.~X.~A. acknowledge the financial support of the Spanish Ministry of Economy and Competitiveness under Grant Consolider nanoTHERM CSD2010-00044, TEC2015-67462-C2-2-R (MINECO/FEDER), TEC2015-67462-C2-1-R (MINECO/FEDER). T.~G.~M.~acknowledges the support of a Ministerio de Ciencia e Innovaci\'on grant MTM2017-82317-P. The authors have been partially funded by the CERCA Programme of the Generalitat de Catalunya.

\begin{appendix}

\section{Calculation of the non-local length and bulk thermal conductivity}
\label{app:kcm}

\begin{figure}[h!]
  \centering
  \subfigure[]{\includegraphics[width=0.49\textwidth]{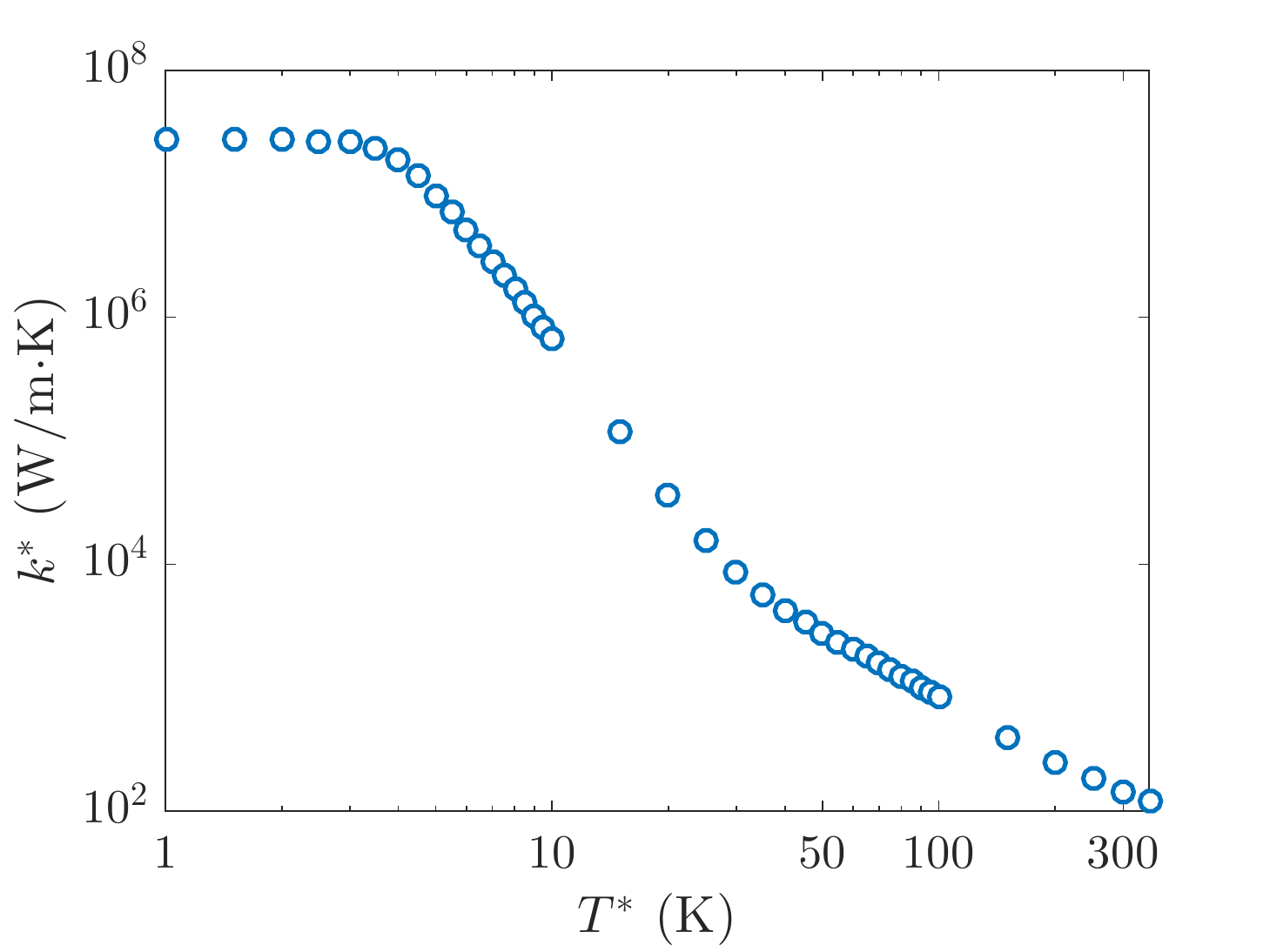}}
  \subfigure[]{\includegraphics[width=0.49\textwidth]{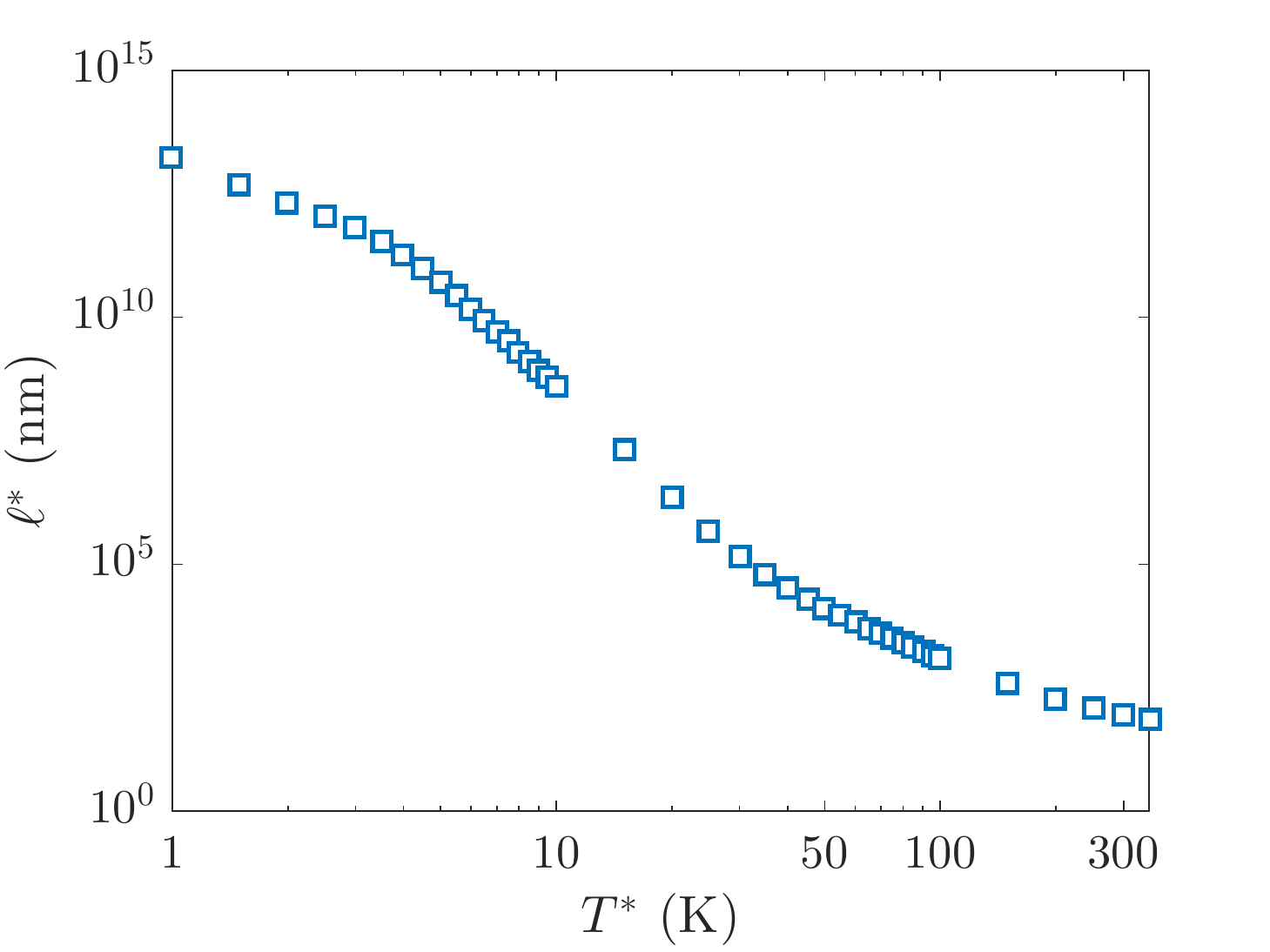}}
  \caption{Bulk values of (a) the thermal conductivity $k^*$ and (b) the non-local length $\ell^*$, both in terms of the temperature $T^*$ obtained from the KCM framework.}
  \label{fig:KCMdata}
\end{figure}

The values of bulk thermal conductivity $k^*(T^*)$ and non-local length $\ell^*(T^*)$, shown in Fig. \ref{fig:KCMdata}, have been obtained in the framework of the KCM. In this model, the thermal conductivity is split in the well-known kinetic regime and a collective contribution emerged from momentum conserving (normal) phonon collisions \cite{Guyer1966a,Torres2017b}. The total thermal conductivity can be expressed thus as an interpolation between the kinetic and collective contributions weighted by a switching factor $\Sigma(T^*)$, which accounts for the relative abundance of normal versus resistive (Umklapp and mass defect) scattering rates:
\begin{equation}
	k^* = \Sigma\cdot k^*_\col + (1-\Sigma)\cdot k^*_\kin.
\end{equation}
In the same line, the non-local length $\ell$, related with the phonon mean free paths, can be expressed as an interpolation through $\Sigma$ between the collective limit derived by Guyer and Krumhansl \cite{Guyer1966b} and the kinetic one derived recently at first order \cite{TorresThesis2017},
\begin{equation}
	\ell^* = \Sigma\cdot \ell^*_\col + (1-\Sigma)\cdot \ell^*_\kin.
\end{equation}
These two magnitudes calculated from first principles are the only ones required to solve the hydrodynamic heat transport equation. For these calculations, an open-source code \cite{Torres2017a} has been used.

\end{appendix}

\bibliographystyle{ieeetr}
\bibliography{ETC}

\begin{thebibliography}{10}

\bibitem{Cahill2003}
D.~G. Cahill, W.~K. Ford, K.~E. Goodson, G.~D. Mahan, A.~Majumdar, H.~J. Maris,
  R.~Merlin, and S.~R. Phillpot, ``Nanoscale thermal transport,'' {\em Journal
  of Applied Physics}, vol.~93, no.~2, pp.~793--818, 2003.

\bibitem{Inyushkin2004}
A.~V. Inyushkin, A.~N. Taldenkov, A.~M. Gibin, A.~V. Gusev, and H.-J. Pohl,
  ``On the isotope effect in thermal conductivity of silicon,'' {\em Physica
  Status Solidi (C)}, vol.~1, no.~11, pp.~2995--2998, 2004.

\bibitem{Li2003}
D.~Li, Y.~Wu, P.~Kim, L.~Shi, P.~Yang, and A.~Majumdar, ``Thermal conductivity
  of individual silicon nanowires,'' {\em Applied Physics Letters}, vol.~83,
  no.~14, pp.~2934--2936, 2003.

\bibitem{Ashenghi1997}
M.~Ashenghi, Y.~K. Leung, S.~S. Wong, and K.~E. Goodson, ``Phonon-boundary
  scattering in thin silicon layers,'' {\em Applied Physics Letters}, vol.~71,
  no.~13, pp.~1798--1800, 1997.

\bibitem{Liu2004}
W.~Liu and M.~Asheghi, ``Phonon–boundary scattering in ultrathin
  single-crystal silicon layers,'' {\em Applied Physics Letters}, vol.~84,
  no.~19, pp.~3819--3821, 2004.

\bibitem{Hoogeboom-Pot2015}
K.~M. Hoogeboom-Pot, J.~N. Hernandez-Charpak, X.~Gu, T.~D. Frazer, E.~H.
  Anderson, W.~Chao, R.~W. Falcone, R.~Yang, M.~M. Murnane, H.~C. Kapteyn, and
  D.~Nardi, ``A new regime of nanoscale thermal transport: {C}ollective
  diffusion increases dissipation efficiency,'' {\em Proceedings of the
  National Academy of Sciences}, vol.~112, no.~16, pp.~4846--4851, 2015.

\bibitem{Johnson2013}
J.~A. Johnson, A.~A. Maznev, J.~Cuffe, J.~K. Eliason, A.~J. Minnich, T.~Kehoe,
  C.~M. Sotomayor~Torres, G.~Chen, and K.~A. Nelson, ``Direct measurement of
  room-temperature nondiffusive thermal transport over micron distances in a
  silicon membrane,'' {\em Physical Review Letters}, vol.~110, no.~2,
  p.~025901, 2013.

\bibitem{Chang2008}
C.-W. Chang, D.~Okawa, H.~Garcia, A.~Majumdar, and A.~Zettl, ``Breakdown of
  {F}ourier's law in nanotube thermal conductors,'' {\em Physical Review
  Letters}, vol.~101, no.~7, p.~075903, 2008.

\bibitem{Callaway1959}
J.~Callaway, ``Model for lattice thermal conductivity at low temperatures,''
  {\em Physical Review}, vol.~113, no.~4, p.~1046, 1959.

\bibitem{Holland1963}
M.~G. Holland, ``Analysis of lattice thermal conductivity,'' {\em Physical
  Review}, vol.~132, no.~6, p.~2461, 1963.

\bibitem{Calvo2018}
M.~Calvo-Schwarzw\"alder, M.~G. Hennessy, P.~Torres, T.~G. Myers, and F.~X.
  Alvarez, ``A slip-based model for the size-dependent effective thermal
  conductivity of nanowires,'' {\em International Communications in Heat and
  Mass Transfer}, vol.~91, pp.~57 -- 63, 2018.

\bibitem{Zhu2017}
C.-Y. Zhu, W.~You, and Z.-Y. Li, ``Nonlocal effects and slip heat flow in
  nanolayers,'' {\em Scientific Reports}, vol.~7, p.~9568, 2017.

\bibitem{Alvarez2009}
F.~X. Alvarez, D.~Jou, and A.~Sellitto, ``Phonon hyrodynamics and
  phonon-boundary scattering in nanosystems,'' {\em Journal of Applied
  Physics}, vol.~105, no.~1, p.~014317, 2009.

\bibitem{Sellitto2012}
A.~Sellitto, F.~X. Alvarez, and D.~Jou, ``Geometrical dependence of thermal
  conductivity in elliptical and rectangular nanowires,'' {\em International
  Journal of Heat and Mass Transfer}, vol.~55, no.~11, pp.~3114--3120, 2012.

\bibitem{Alvarez2007}
F.~X. Alvarez and D.~Jou, ``Memory and nonlocal effects in heat transport: From
  diffusive to ballistic regimes,'' {\em Applied Physics Letters}, vol.~90,
  no.~8, p.~083109, 2007.

\bibitem{Ma2012}
Y.~Ma, ``Size-dependent thermal conductivity in nanosystems based on
  non-{F}ourier heat transfer,'' {\em Applied Physics Letters}, vol.~101,
  p.~211905, 2012.

\bibitem{Dong2014}
Y.~Dong, B.-Y. Cao, and Z.-Y. Guo, ``Size dependent thermal conductivity of si
  nanosystems based on phonon gas dynamics,'' {\em Physica E: Low-dimensional
  Systems and Nanostructures}, vol.~56, pp.~256--262, 2014.

\bibitem{Tzou2011}
D.~Y. Tzou, ``Nonlocal behaviour in phonon transport,'' {\em International
  Journal of Heat and Mass Transfer}, vol.~54, no.~1, pp.~475--481, 2011.

\bibitem{Guyer1966a}
R.~A. Guyer and J.~A. Krumhansl, ``{Solution of the Linearized Phonon Boltzmann
  Equation},'' {\em Physical Review}, vol.~148, no.~2, p.~766, 1966.

\bibitem{Guyer1966b}
R.~A. Guyer and J.~A. Krumhansl, ``{Thermal Conductivity, Second Sound, and
  Phonon Hydrodynamic Phenomena in Nonmetallic Crystals},'' {\em Physical
  Review}, vol.~148, no.~2, p.~778, 1966.

\bibitem{Guo2015}
Y.~Guo and M.~Wang, ``Phonon hydrodynamics and its applications in nanoscale
  heat transport,'' {\em Physics Reports}, vol.~595, pp.~1--44, 2015.

\bibitem{Guo2018}
Y.~Guo and M.~Wang, ``Phonon hydrodynamics for nanoscale heat transport at
  ordinary temperatures,'' {\em Physical Review B}, vol.~97, p.~035421, Jan
  2018.

\bibitem{Lee2015}
S.~Lee, D.~Broido, K.~Esfarjani, and G.~Chen, ``{Hydrodynamic phonon transport
  in suspended graphene},'' {\em Nature Communications}, vol.~6, p.~6290, 2015.

\bibitem{Ziabari2018}
A.~Ziabari, P.~Torres, B.~Vermeersch, Y.~Xuan, X.~Cartoix{\`{a}},
  A.~Torell{\'{o}}, J.-H. Bahk, Y.~R. Koh, M.~Parsa, P.~D. Ye, F.~X. Alvarez,
  and A.~Shakouri, ``{Full-field thermal imaging of quasiballistic crosstalk
  reduction in nanoscale devices},'' {\em Nature Communications}, vol.~9,
  no.~1, p.~255, 2018.

\bibitem{Dong2011}
Y.~Dong, B.-Y. Cao, and Z.-Y. Guo, ``Generalized heat conduction laws based on
  thermomass theory and phonon hydrodynamics,'' {\em Journal of Applied
  Physics}, vol.~110, no.~6, p.~063504, 2011.

\bibitem{Majumdar1993}
A.~Majumdar, ``Microscale heat conduction in dielectric films,'' {\em Journal
  of Heat Transfer}, vol.~115, no.~1, pp.~7--16, 1993.

\bibitem{Boltzmann1872}
L.~Boltzmann, ``{Weitere Studien \"uber das W\"armegleichgewicht unter
  Gasmolek\"ulen},'' {\em Sitzungsberichte der Akademie der Wissenschaften
  Wien}, vol.~66, no.~II, pp.~275--370, 1872.

\bibitem{Jou1996}
D.~Jou, J.~Casas-Vazquez, and G.~Lebon, {\em {Extended Irreversible
  Thermodynamics}}.
\newblock Springer, 2nd~ed., 1996.

\bibitem{Xu2014}
M.~Xu, ``Slip boundary condition of heat flux in knudsen layers,'' {\em
  Proceedings of the Royal Society of London A: Mathematical, Physical and
  Engineering Sciences}, vol.~470, no.~2161, 2014.

\bibitem{Torres2017b}
P.~Torres, A.~Torell\'o, J.~Bafaluy, J.~Camacho, X.~Cartoix\`a, and F.~X.
  Alvarez, ``{First principles Kinetic-Collective thermal conductivity of
  semiconductors},'' {\em Physical Review B}, vol.~95, no.~4, p.~165407, 2017.

\bibitem{TorresThesis2017}
P.~Torres, {\em Thermal transport in semiconductors: first principles and
  phonon hydrodynamics}.
\newblock PhD thesis, Universitat Aut\`onoma de Barcelona, 2017.

\bibitem{Sellitto2015}
A.~Sellitto, I.~Carlomagno, and D.~Jou, ``Two-dimensional phonon hydrodynamics
  in narrow strips,'' {\em Proceedings of the Royal Society of London A:
  Mathematical, Physical and Engineering Sciences}, vol.~471, no.~2182, 2015.

\bibitem{Sellitto2010}
A.~Sellitto, F.~X. Alvarez, and D.~Jou, ``Temperature dependence of boundary
  conditions in phonon hydrodynamics of smooth and rough nanowires,'' {\em
  Journal of Applied Physics}, vol.~107, no.~11, p.~114312, 2010.

\bibitem{Ziman2001}
J.~M. Ziman, {\em Electrons and {P}honons: {T}he {T}heory of {T}ransport
  {P}henomena in {S}olids}.
\newblock Oxford University Press, 2001.

\bibitem{Ockendon1995}
H.~Ockendon and J.~R. Ockendon, {\em {Viscous flow}}.
\newblock Cambridge University Press, 1995.

\bibitem{ebert1965}
W.~A. Ebert and E.~M. Sparrow, ``Slip flow in rectangular and annular ducts,''
  {\em Journal of Basic Engineering}, vol.~87, no.~4, pp.~1018--1024, 1965.

\bibitem{berman1955}
R.~Berman, E.~L. Foster, and J.~M. Ziman, ``Thermal conduction in artificial
  sapphire crystals at low temperatures i. nearly perfect crystals,'' {\em
  Proceedings of the Royal Society of London A: Mathematical, Physical and
  Engineering Sciences}, vol.~231, no.~1184, pp.~130--144, 1955.

\bibitem{Torres2017a}
P.~Torres, {\em Kinetic Collective Model: BTE-based hydrodynamic model for
  thermal transport}, 2017 (accessed November 9, 2017).

\bibitem{Zhang2007}
Z.~M. Zhang, {\em {Nano/Microscale Heat Transfer}}.
\newblock McGraw Hill Professional, 2007.

\bibitem{Asheghi1998}
M.~Asheghi, M.~Touzelbaev, K.~Goodson, Y.~Leung, and S.~Wong,
  ``Temperature-dependent thermal conductivity of single-crystal silicon layers
  in {SOI} substrates,'' {\em Journal of Heat Transfer}, vol.~120, no.~1,
  pp.~30--36, 1998.

\bibitem{Vermeersch2015}
B.~Vermeersch, J.~Carrete, N.~Mingo, and A.~Shakouri, ``{Superdiffusive heat
  conduction in semiconductor alloys. I. Theoretical foundations},'' {\em
  Physical Review B}, vol.~91, p.~085202, 2015.

\end{thebibliography}

\end{document}